\documentclass[longauth]{aa} 
\pdfoutput=1 
\usepackage[utf8]{inputenc}
\usepackage[T1]{fontenc}
\usepackage{color}
\usepackage{multirow}
\usepackage{graphicx}
\usepackage{txfonts}
\usepackage[hidelinks]{hyperref}
\usepackage{xcolor}
\hypersetup{
    colorlinks,
  linkcolor={red!50!black},
    citecolor={blue!50!black},
   urlcolor={blue!80!black}
}
\usepackage{gensymb}
\usepackage{wasysym}
\usepackage[flushleft]{threeparttable}

\usepackage{natbib}

\usepackage{footmisc}

\begin{document} 
	
\title{Activity and Magnetic Field Structure of the Sun-Like Planet Hosting Star HD 1237}
	
\author{J. D. Alvarado-G\'omez\inst{1,2}, G. A. J. Hussain\inst{1,4}, J. Grunhut\inst{1}, R. Fares\inst{3}, J.-F. Donati\inst{4,5}, E. Alecian\inst{6,7}, O. Kochukhov\inst{8}, M. Oksala\inst{7}, J. Morin\inst{9}, S. Redfield\inst{10}, O. Cohen\inst{11}, J. J. Drake\inst{11}, M. Jardine\inst{3}, S. Matt\inst{12},  P. Petit\inst{4,5}, 
	\and	
	F. M. Walter\inst{13}
          }
          \institute{\inst{1} European Southern Observatory,
              Karl-Schwarzschild-Str. 2, 85748 Garching bei M\"unchen, Germany\\
              \email{jalvarad@eso.org} \\
              \inst{2} Universit\"ats-Sternwarte M\"unchen, Ludwig-Maximilians-Universit\"at, Scheinerstr.~1, 81679 M\"unchen, Germany \\
              \inst{3} INAF-Osservatorio Astrofisico di Catania, Via Santa Sofia 87, 95123 Catania, Italy \\
              \inst{4} Institut de Recherche en Astrophysique et Plan\'etologie, Universit\'e de Toulouse, UPS-OMP, F-31400 Toulouse, France \\
              \inst{5} CNRS, Institut de Recherche en Astrophysique et Plan\'etologie, 14 Avenue Edouard Belin, F-31400 Toulouse, France \\
              \inst{6} UJF Grenoble 1 CNRS-INSU, Institut de Plan\'etologie et d'Astrophysique de Grenoble (IPAG) UMR 5274, France \\
              \inst{7} LESIA, Observatoire de Paris, CNRS UMR 8109, UPMC, Universit\'{e} Paris Diderot, 5 place Jules Janssen, 92190, Meudon, France \\
              \inst{8} Department of Physics and Astronomy, Uppsala University, BOX 516, SE-751 20 Uppsala, Sweden \\
              \inst{9} LUPM-UMR 5299, CNRS \& Universit\'e Montpellier, Place Eug\'ene Bataillon, 34095 Montpellier Cedex 05, France \\
              \inst{10} Astronomy Department, Van Vleck Observatory, Wesleyan University, 96 Foss Hill Drive, Middletown, CT 06459, USA \\
              \inst{11} Harvard-Smithsonian Center for Astrophysics, 60 Garden Street, Cambridge, MA 02138, USA \\ 
              \inst{12} Department of Physics and Astronomy, University of Exeter, Stocker Road, Exeter EX4 4SB, UK \\
              \inst{13} Department of Physics and Astronomy, Stony Brook University, Stony Brook NY 11794-3800, USA
             }
             
   \date{Received -----; accepted -----}

 
\abstract{We analyse the magnetic activity characteristics of the planet hosting Sun-like star, HD 1237, using HARPS spectro-polarimetric time-series data. We find evidence of rotational modulation of the magnetic longitudinal field measurements consistent with our ZDI analysis, with a period of 7 days. We investigate the effect of customising the LSD mask to the line depths of the observed spectrum and find that it has a minimal effect on shape of the extracted Stokes V profile but does result in a small increase in the S/N ($\sim$ 7\%). We find that using a Milne-Eddington solution to describe the local line profile provides a better fit to the LSD profiles in this slowly rotating star, which also impacts the recovered ZDI field distribution. We also introduce a fit-stopping criterion based on the information content (entropy) of the ZDI maps solution set. The recovered magnetic field maps show a strong (+90 G) ring-like azimuthal field distribution and a complex radial field dominating at mid latitudes ($\sim$45 degrees). Similar magnetic field maps are recovered from data acquired five months apart. Future work will investigate how this surface magnetic field distribution impacts the coronal magnetic field and extended environment around this planet-hosting star.}   


\titlerunning{Activity and Magnetic Field Structure of HD 1237}
\authorrunning{Alvarado-G\'omez et al.}
\maketitle
%

\section{Introduction}\label{sec_intro}

\noindent Observational studies of magnetism in late-type stars have evolved dramatically during the last two decades; from the classical chromospheric activity diagnostics (e.g. Mount Wilson H-K project, \citealt{1995ApJ...438..269B}) to spectro-polarimetric snapshot surveys (the BCool project, \citealt{2014MNRAS.444.3517M}) and detailed long-term magnetic monitoring (e.g. \citealt{2012A&A...540A.138M}). This has been enabled by the advent of improved instrumentation (e.g. ESPaDOnS@CFHT, \citealt{2003ASPC..307...41D}) together with advanced data analysis techniques for detection (e.g. Least Squares Deconvolution, \citealt{1997MNRAS.291..658D}) and mapping (e.g. Zeeman Doppler Imaging, \citealt{1997A&A...326.1135D}) of magnetic fields. These recent studies have opened new possibilities for different areas of astrophysical research, in particular, on dynamo processes and the origin of stellar magnetic fields across the HR diagram (see \citealt{2009ARA&A..47..333D}). In the case of main sequence solar-type stars, complex and relatively weak ($\ll$ 1 kG) large-scale surface field topologies have been reported. The appearance of ring-like structures of significant (and even dominant) toroidal fields in these stars seems to be connected with the rotation period, and therefore with the dynamo mechanism behind their generation \citep{2008MNRAS.388...80P}. 

From the perspective of exoplanet studies, activity-related signatures in late-type stars are known to affect the detection techniques in the form of radial-velocity jitter and photometric flicker \citep{2014AJ....147...29B}. By simulating the effects induced by active regions and spots in Sun-like stars, recent tools have been developed to estimate and remove their contribution from the observations (e.g. \citealt{2014ApJ...796..132D}). Other  studies have considered the detectability of planets around active cool stars, by modelling the stellar activity from recovered starspot and magnetic field maps \citep{2014MNRAS.438.2717J, 2014MNRAS.444.3220D}. It is clear that knowing the characteristics of the stellar magnetic field is crucial for addressing the presence of exo-planets in a given system (e.g. the case of $\epsilon$ Eridani, \citealt{2012ApJS..200...15A, 2014A&A...569A..79J}). 

In addition, the stellar magnetic field dominates the environment around late-type stars. This includes transient events such as flares and coronal mass ejections \citep{2011LRSP....8....6S}, and the development of persistent solar-like winds and astrospheres \citep{2004LRSP....1....2W}. These phenomena are known to have a profound impact on the structure of exoplanet atmospheres, a critical factor in the habitability of exo-planetary systems (see \citealt{2013A&A...557A..67V, 2011ApJ...738..166C, 2014ApJ...790...57C}). They can erode atmospheres through thermal evaporation, or non-thermal processes, such as sputtering and ion pick-up. Significant mass loss has been detected for exoplanets that is driven by the stellar wind \citep{2003Natur.422..143V, 2010ApJ...717.1291L, 2012ApJ...751...86J}. Models of this interaction require accurate knowledge of the wind properties \citep{2010ApJ...709..670E, 2011A&A...532A...6S}, and therefore, of the host-star surface magnetic field. However, given the observational limitations, robust surface field distributions are known for a very limited number of Sun-like planet-hosting stars (e.g. \citealt{2013MNRAS.435.1451F, 2014IAUS..302..180F}). In this context, detailed studies of these systems are very valuable resources not only as direct stellar counterparts of our solar system, but also for the growing interest in finding suitable Earth-like life supporting places in the Universe.

In this article we present the detailed study of one such planet-hosting Sun-like star (HD 1237), in which we investigate the large scale magnetic field and chromospheric activity. This is the first step in characterising the impact the stellar magnetic field on the circumstellar environment around this system. In particular, the conditions and possible interactions via the magnetically-driven stellar wind, with the Jupiter-size exoplanet which comes as close as 0.25 AU in its orbit \citep{2001A&A...375..205N}. In section \ref{sec_properties}, we summarise the main properties of the star. Details of the observations and calibration procedures are given in section \ref{sec_data}. We present the activity diagnostics and variability in section \ref{sec_activity}. Section \ref{sec_mag_signatures} contains a description of the implemented technique for extracting the magnetic field signatures from the spectro-polarimetric data. The required steps for the imaging procedure and the resulting surface field maps are presented in section \ref{sec_zdi}. In section \ref{sec_summary}, we discuss our findings in context of previous and ongoing studies of solar-type stars. Our main conclusions are summarised in section \ref{sec_conclusions}.    

\section{HD 1237 stellar properties}\label{sec_properties}

\noindent HD 1237 (GJ 3021) is a bright ($V_{\rm mag}$ = 6.58), Sun-like star (G8V) located about 17.5 pc from the Sun, in the southern constellation of Hydrus \citep{2010MNRAS.403.1949K}. This object is a relatively young ($\sim$\,0.88 Gyr), chromospherically active, and confirmed exo-planet host star. \cite{2001A&A...375..205N} used the enhanced chromospheric activity to explain the large residuals arising from the best Keplerian orbital solution of the planet ($M_{\rm p}\sin(i) = 3.37 \pm 0.14$ M$_{\jupiter}$, $P_{\rm orb} = 133.7 \pm 0.2$ days, $e = 0.51 \pm 0.02$,  $a = 0.49$ AU). 

Table \ref{tab_1} contains the basic stellar properties of HD 1237 taken from \cite{2010ApJ...720.1290G}, \cite{2006A&A...460..695T} and \cite{2005A&A...443..609S}. Rotation period ($P_{\rm rot}$) estimates are sparse, ranging from $\sim$\,4.0 to 12.6 days, as summarized by \citep{2010MNRAS.408.1606W}. As presented in section \ref{sec_parameters}, we obtain $P_{\rm rot} = 7.0 \pm 0.7$ from our observations (Sect. \ref{sec_data}). In addition, we estimated the radial velocity, $v_{\rm R}$ and the rotational velocity, $v\sin i$, using an automatic spectral classification tool (MagIcS, \citealt{2012MNRAS.425.2948D}) and the fundamental properties of the star (Table \ref{tab_1}). Using several of our observed spectra, we found on average a $v\sin i$ of $5.3 \pm 1.0$ km s$^{-1}$ and a $v_{\rm R}$ of $- 5.2 \pm 0.2$ km s$^{-1}$. Literature values for $v\sin i$ range between $\sim$ 4.5 to 5.5 km s$^{-1}$ \citep{2001A&A...375..205N, 2006A&A...460..695T, 2009A&A...493.1099S}. However, as indicated by \cite{2001A&A...375..205N}, the $v\sin i$ value can be over-estimated for metal-rich stars such as HD 1237. For the subsequent analysis we adopted a $v\sin i$ of $5.3$ km s$^{-1}$ which is consistent, within the errors, with the value reported by \citet{2001A&A...375..205N} and \citet{2006A&A...460..695T}.


\begin{table}[h]
\caption{HD 1237 basic properties.}             
\label{tab_1}      
\begin{threeparttable}
\centering                          
\begin{tabular}{l c c}        
\hline\hline                 
Parameter & Value & Reference \\
\hline
S. Type & G8V & \protect{\cite{2006A&A...460..695T}}\\ 
$T_{\rm eff}$ [K] & $5572 \pm 40$ & \protect{\cite{2010ApJ...720.1290G}} \\
$\log(g)$ & $4.58 \pm 0.2$ & \protect{\cite{2010ApJ...720.1290G}} \\
$R_{*}$ [R$_{\odot}$] & $0.86 \pm 0.07$ & \protect{\cite{2010ApJ...720.1290G}} \\
$M_{*}$ [M$_{\odot}$] & $1.0 \pm 0.1$ & \protect{\cite{2010ApJ...720.1290G}} \\
$v\sin i$ [km s$^{-1}$]$^\dagger$ & $5.3 \pm 1.0$  & This work \\
$v_{\rm R}$ [km s$^{-1}$] & $-5.2 \pm 0.2$ & This work \\ 
$P_{\rm rot}$ [days] & $7.0 \pm 0.7$ & This work \\
$\log(L_{\rm X})$ & $29.02 \pm 0.06$ & \protect{\cite{2008ApJ...687.1339K}}\\
Age [Gyr]$^\ddagger$ & $\sim0.88$ & \protect{\cite{2005A&A...443..609S}} \\
\hline                                   
\end{tabular}
\begin{tablenotes}
\item {\small $\dagger$: Other reports include 4.5 km s$^{-1}$ \citep{2009A&A...493.1099S}, $5.1 \pm 1.2$ km s$^{-1}$ \citep{2006A&A...460..695T} and $5.5 \pm 1.0$ km s$^{-1}$ \citep{2001A&A...375..205N}.}
\item {\small $\ddagger$: Age estimates range from 0.15 to 0.88 Gyr using various methods (see \citealt{2001A&A...375..205N} and \citealt{2005A&A...443..609S}). 0.88 Gyr corresponds to the age determined using isochrones.}
\end{tablenotes}
\end{threeparttable}
\end{table}

\section{Observational data}\label{sec_data}

\noindent We obtained observations using the polarimetric mode \citep{2011Msngr.143....7P} of the HARPS echelle spectrograph \citep{2003Msngr.114...20M} at the ESO 3.6\,m telescope at La Silla Observatory. The wavelength coverage of the observations range from 378 nm to 691 nm, with a 8 nm gap starting at 526 nm. 

Data were reduced using the REDUCE package \citep{2002A&A...385.1095P, 2011A&A...525A..97M}, which was modified for the HARPS instrument configuration. This package produces an optimal extraction of the bias-subtracted spectra after flat-fielding corrections and cosmic ray removal have been carried out. The continuum level is determined by masking out the strongest, broadest features (e.g. the Balmer lines) and then fitting a smooth slowly varying function to the envelope of the entire spectrum. Spectra are obtained with resolutions varying from 95\,000 to 113\,000, depending on the wavelength, with a median value of 106\,000. Uncertainties are derived for each pixel assuming photon statistics. The star was observed at two epochs separated by 5 months (July and December) in 2012. A summary of the observations is presented in Table \ref{tab_2}.

\begin{table}[h]
\caption{Journal of observations. The columns contain the date, the corresponding Heliocentric Julian Date (HJD), the start time of the observations in UT, the exposure times, and the Stokes I peak Signal-to-Noise ratio (S/N). The rotational phase ($\Phi$) listed in the last column is calculated using the rotation period derived in this work ($P_{\rm rot} = 7.0$ d).}             
\label{tab_2}      
\begin{threeparttable}
\centering                          
{\small 
\begin{tabular}{l c c c c c}        
\hline\hline                 
Date & HJD & UT & $t_{\rm exp}$ & Stokes I & Phase \\    
(2012) & (2400000+) & & [s] & Peak S/N & ($\Phi$) \\    
\hline                        
\multicolumn{6}{l}{\textit{First epoch}}\\
Jul 15 & 56123.359 & 08:04:36 & 3600.0 & 955 & 0.000\\      
Jul 16 & 56124.442 & 10:03:43 & 3600.0 & 1214 & 0.155\\
Jul 17 & 56125.399 & 09:01:10 & 3600.0 & 841 & 0.291\\
Jul 18 & 56126.361 & 08:06:35 & 3600.0 & 877 & 0.429\\
Jul 19 & 56127.440 & 10:01:33 & 3600.0 & 753 & 0.583\\
Jul 20 & 56128.356 & 08:00:16 & 3600.0 & 1098 & 0.714\\
Jul 21 & 56129.374 & 08:25:27 & 3600.0 & 911 & 0.859\\
Jul 22 & 56130.438 & 09:47:58 & 4800.0 & 1092 & 1.011\\
Jul 23$^\dagger$ & 56131.397 & 08:32:33 & 6680.0 & 660 & 1.148\\
Jul 31 & 56139.314 & 06:49:52 & 4800.0 & 926 & 2.280\\
Aug 02 & 56141.303 & 06:32:21 & 5000.0 & 1040 & 2.564\\
\multicolumn{6}{l}{\textit{Second epoch [20.29+ rotation cycles since Jul 15 2012]}}\\
Dec 04 & 56265.045 & 00:35:27 & 2800.0 & 1205 & 0.000\\
Dec 05 & 56266.045 & 00:35:21 & 2800.0 & 964 & 0.142\\
Dec 06 & 56267.044 & 00:34:40 & 2800.0 & 1018 & 0.285\\
Dec 07 & 56268.044 & 00:33:59 & 2800.0 & 722 & 0.428\\
\hline                                   
\end{tabular}}
\begin{tablenotes}
\item {\small $\dagger$: The listed values correspond to two spectro-polarimetric exposures merged in a single observation due to bad weather conditions.}
\end{tablenotes}
\end{threeparttable}
\end{table}

\noindent  The exposure times listed correspond to one circularly polarised spectrum (Stokes V) which results from combining four individual sub-exposures using the ratio method. As is explained in \citet{1997MNRAS.291..658D}, the polarization signal is obtained by dividing spectra with perpendicular (orthogonal) polarization states (for HARPSpol and Stokes V: $45^{\circ}$, $135^{\circ}$, $225^{\circ}$ and $315^{\circ}$, using the quarter waveplate). Additionally, a null-polarisation spectrum is constructed in order to check for possible spurious polarisation contributions in the observations. More details can be found in \citet{2009PASP..121..993B}. Due to bad weather conditions, two consecutive Stokes V spectra were added together for the night of 2012 Jul 23.

\section{Magnetic activity and variability}\label{sec_activity}

In order to characterise the chromospheric activity level of the star during the observed epochs, we use the CaII H (396.8492 nm) \& K (393.3682 nm) lines and the classic Mount Wilson S-index, $S_{\rm MW}$, defined as

\begin{equation}\label{eq_1}
S_{\rm MW} = \dfrac{H + K}{R + V}\mbox{ .}
\end{equation}

\noindent $H$ and $K$ represent the fluxes measured in each of the Ca line cores using $0.105$ nm wide spectral windows. $R$ and $V$ are the fluxes measured in the continuum over $2$ nm windows centred at $390.1$ nm and $400.1$ nm respectively, on both sides of the CaII region. 

\subsection{Index calibration}

To compare the activity level of HD 1237 with other stars, we need to convert the measured HARPS S-index, $S_{\rm H}$, to the Mount Wilson scale. For this, we require a calibration factor, $\alpha$, which is an instrument-dependent quantity that linearly relates the values for the classic $S_{\rm MW}$ and the HARPS fluxes $H$, $K$, $R$, $V$:

\begin{equation}\label{eq_2}
S_{\rm MW} = \underbrace{\alpha\left(\dfrac{H + K}{R + V}\right)_{\rm H}}_\textit{\normalsize{S$_{\rm H}$}}
\end{equation}

\begin{figure}[!h]
\centering 
\includegraphics[trim=1.45cm 0.45cm 0.2cm 1.0cm, clip=true, width=\hsize]{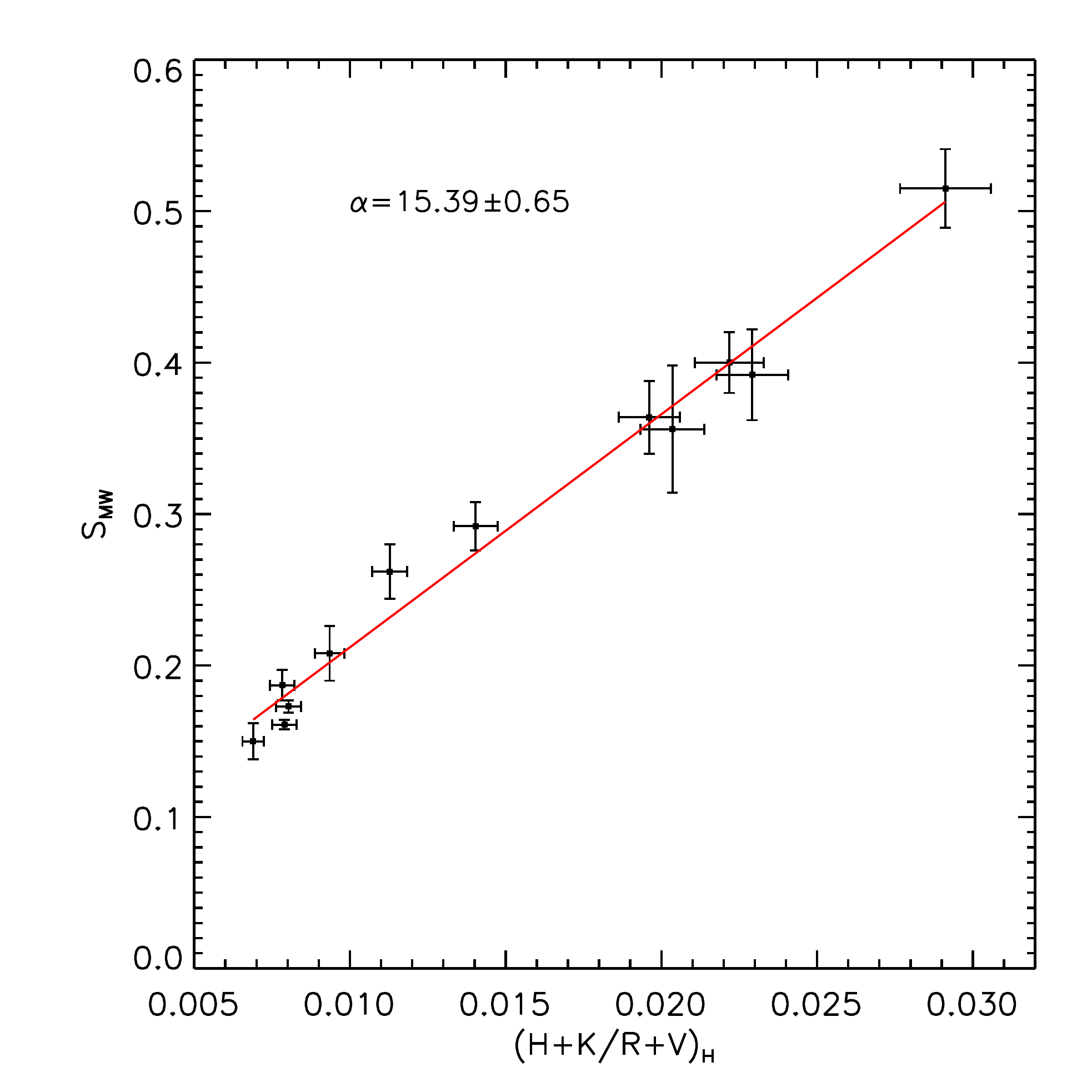}
\caption{Linear fit between \cite{2000A&A...361..265S} $S_{\rm MW}$ values and the HARPS fluxes $\left(\frac{H+K}{R+V}\right)_{\rm H}$. A 5\% error is estimated in our HARPS flux measurements.}
\label{fig_2}
\end{figure}

\noindent We estimated $\alpha$ by including a set of stars with previous measurements of chromospheric activity via $S_{\rm MW}$ \citep{2000A&A...361..265S} and within the HARPS observations database. The spectral type of the reference stars and their reported $S_{\rm MW}$ values are listed in the Table \ref{tab_3}. The linear relation between $S_{\rm MW}$ and the HARPS fluxes is plotted in the Figure \ref{fig_2}. The derived calibration factor, within the 1-$\sigma$ uncertainty, is 

\begin{equation}\label{eq_3}
\alpha = 15.39 \pm 0.65\mbox{.}
\end{equation}

\noindent All spectra were co-aligned using a high S/N HARPS solar spectrum as reference. We estimate a 5\% typical error size in our HARPS flux measurements based on the possible differences in the continuum normalisation, which was performed in the same way for all the stars in the calibration. Therefore, the errors in $S_{\rm H}$ are dominated by the conversion procedure.

\begin{table}[h]
\caption{Stars included in the $\alpha$ calibration.}             
\centering                          
\begin{tabular}{l l c c}        
\hline\hline                 
Name & S. Type & $S_{\rm MW}$ & $\sigma_{\rm MW}$ \\    
\hline                        
HD1835 & G3V & $0.364$ & $0.024$ \\      
HD10700 & G8.5V & $0.173$ & $0.004$ \\
HD22049 & K2Vk & $0.515$ & $0.026$ \\
HD23249 & K1III\,-\,IV & $0.150$ & $0.012$ \\
HD26965 & G9III\,-\,IV & $0.208$ & $0.018$ \\
HD30495 & G1.5V & $0.292$ & $0.016$ \\
HD61421 & F5IV\,-\,V & $0.187$ & $0.010$ \\
HD76151 & G3V & $0.262$ & $0.018$ \\
HD115617 & G7V & $0.161$ & $0.003$ \\
HD149661 & K2V & $0.356$ & $0.042$ \\
HD152391 & G8.5Vk & $0.392$ & $0.030$ \\
HD155885 & K1V & $0.400$ & $0.020$ \\
\hline                                   
\end{tabular}
\label{tab_3}      
\end{table}

\noindent We proceed with the estimation of the activity index $S_{\rm H}$, with the corresponding indicators $R_{\rm HK}$ \citep{1982A&A...107...31M} and $R^\prime_{\rm HK}$ \citep{1984ApJ...279..763N}, which account for colour and photospheric correction respectively.

\subsection{Activity indicators}\label{sec_S_index}

\begin{figure*}[!ht]
\centering
\includegraphics[width=\hsize/2]{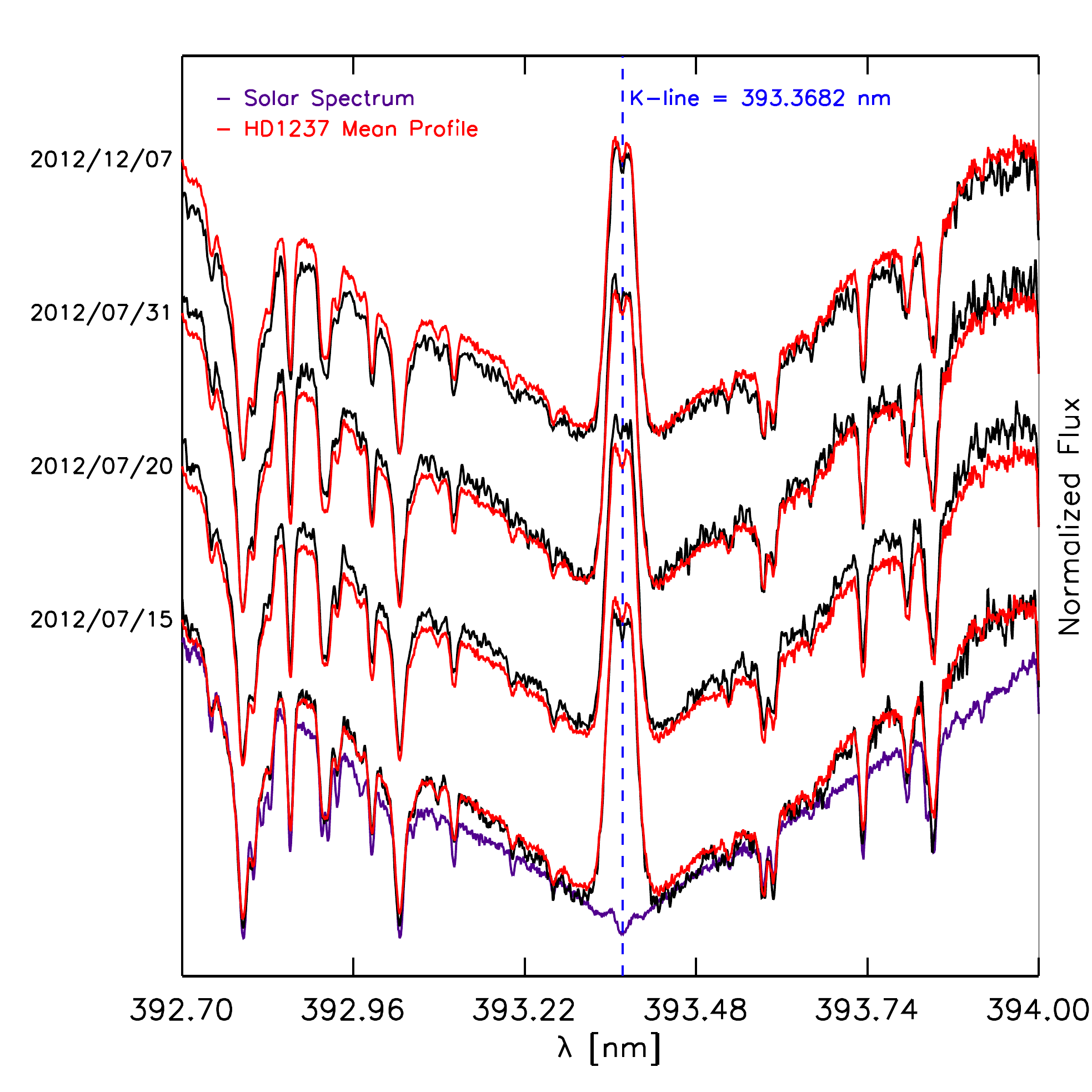}\includegraphics[width=\hsize/2]{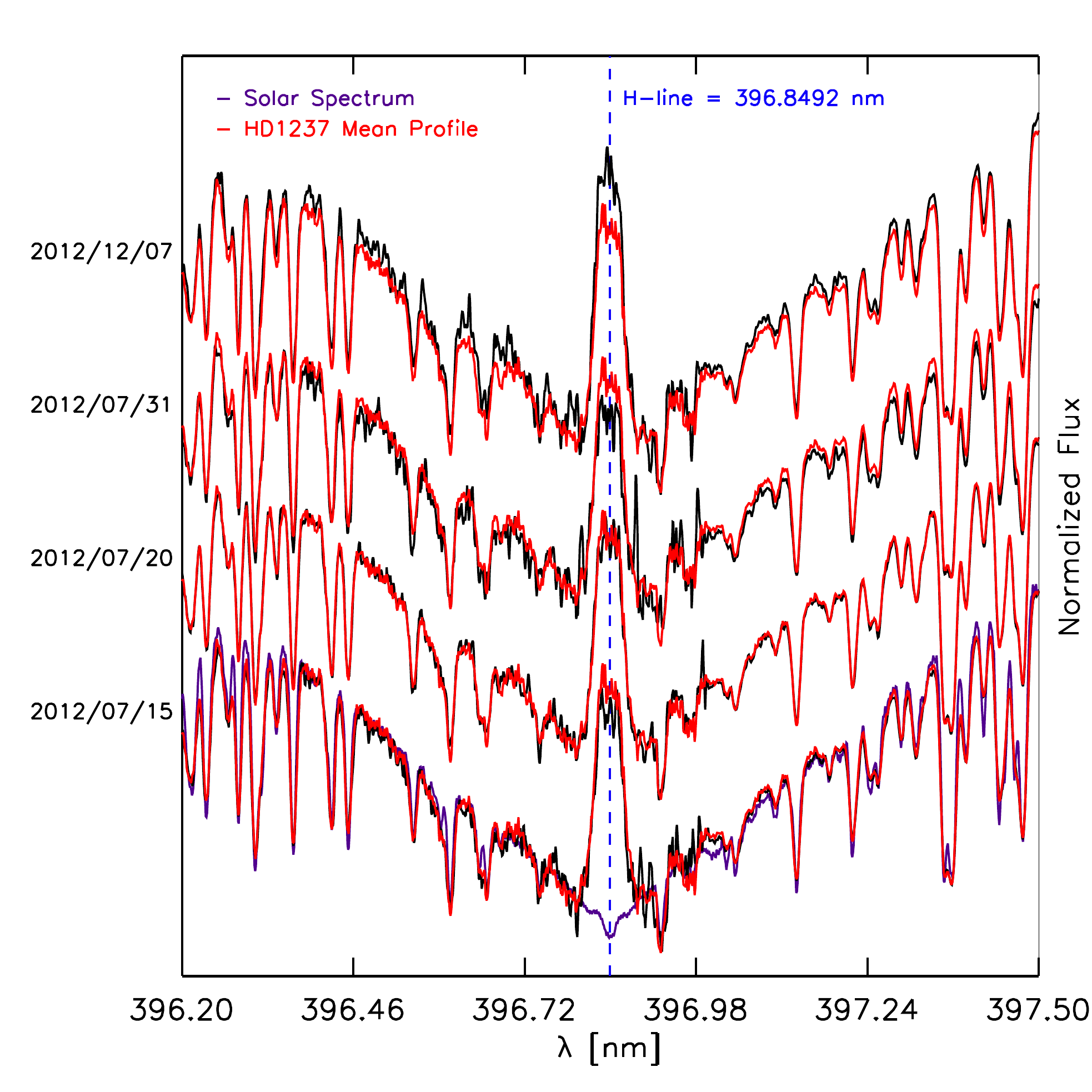}
\caption{Core regions of the CaII K (left) and H (right) lines of HD 1237. Four spectra from our sample are plotted, with the corresponding dates on the left y-axis and vertically shifted (0.25 units) for visualisation purposes. The black spectrum at the top was taken at a much later epoch (2012 Dec 07). The red line shows the mean profile for the entire dataset, while the purple line is a HARPS solar spectrum used as reference.}
\label{fig_3}
\end{figure*}

\noindent Figure \ref{fig_3} shows the cores of the CaII H and K lines in the HARPS normalized spectra of HD 1237 for different observations, which are compared with the mean profile (red) derived from the entire data set. For the averaging procedure we take into account the slight differences in the wavelength range from each observation by an interpolation procedure to match the largest wavelength data points in the observed spectra. A high S/N HARPS Solar spectrum\footnote[2]{S/N: 347 @550 nm, Date: 2007 Apr 12 -- Low activity period.}, is shown in purple as reference. 

The upper plot shows one observation at a later epoch (2012 Dec 07), where it is possible to observe a variation in the line profile. This is interpreted as a slight change in the chromospheric/photospheric activity of the star in comparison with the mean behaviour of the red line (that is dominated by profiles from the first epoch), especially in the H line (Fig. \ref{fig_3}, right panel). The K\,-\,line region of the spectrum contained more noise. No change in the activity level of the star is visible in this particular line. 

\begin{figure*}[ht]
\centering 
\includegraphics[trim=1.2cm 0.8cm 1.8cm 1.4cm, clip=true, width=\hsize]{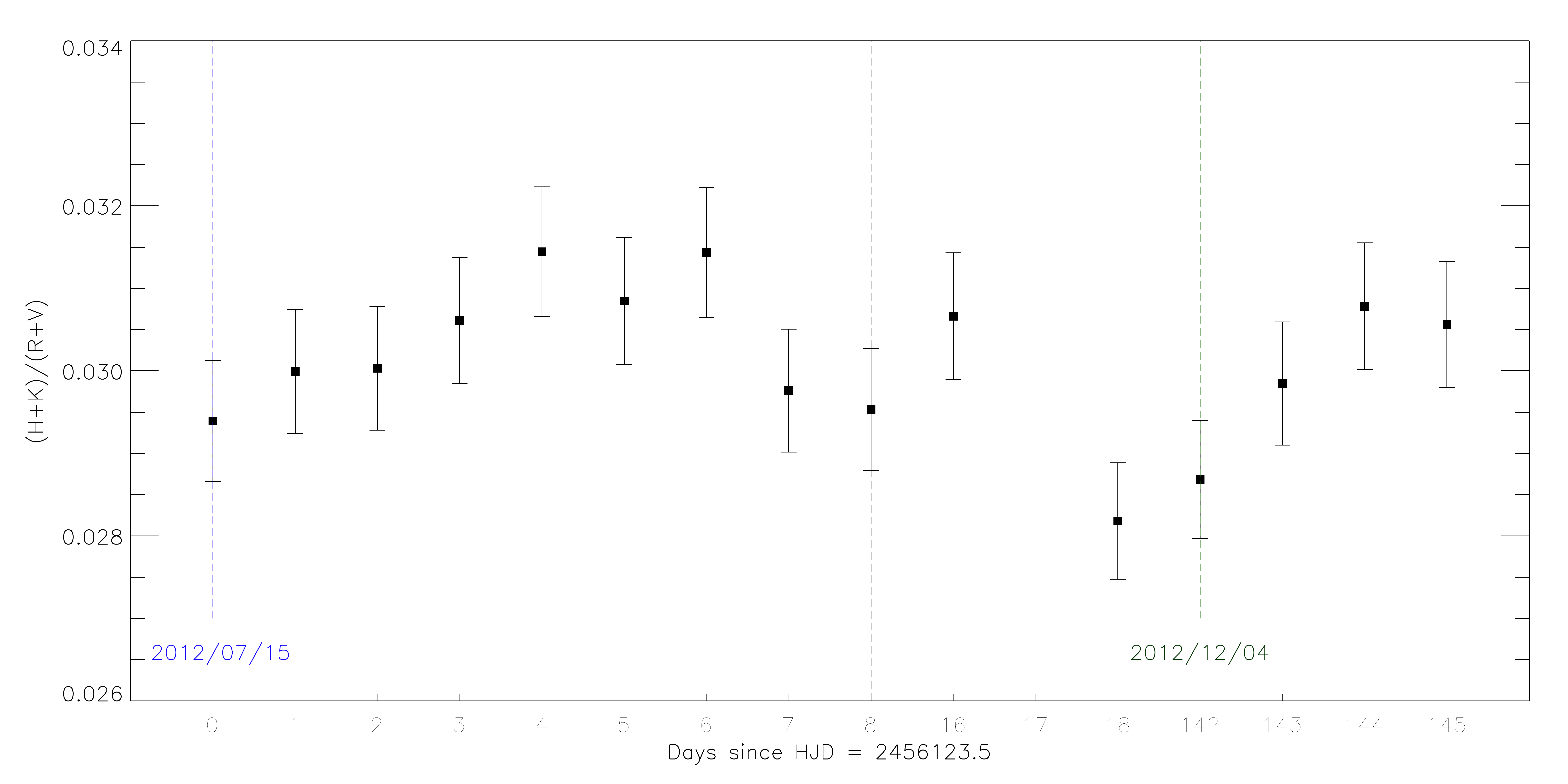} 
\caption{HARPS fluxes, $(H+K)$/$(R+V)$, for the available observations of HD 1237. The blue and green vertical lines denote the beginning of each observed epoch. The x-axis contains the number of days since the Heliocentric Julian Date (HJD = 2456123.5) of the first observation (2012 Jul 15). The last two data points of the first epoch are unevenly distributed (after the vertical black line). We estimate a 5\% error in our measurements.}
\label{fig_4}
\end{figure*}

Figure \ref{fig_4} shows the measured HARPS fluxes, $(H+K)$/$(R+V)$, for each observation starting from 2012 Jul 15 (vertical blue line, HJD = 2456123.5). The x-axis units correspond to days after this initial date. The activity of the star showed a marginal variation, in a similar way as in the first epoch. This may be due to the rotation of active regions over the stellar surface. The similarities between the activity levels between both epochs, could be an indication of a stable large-scale magnetic field configuration. 

The chromospheric activity level of the star can be quantified by using Eq. (\ref{eq_2}), and the derived value for $\alpha$ in Eq. (\ref{eq_3}). We estimate an average value of $S_{\rm H} \simeq 0.46 \pm 0.02$ for the available observations of HD 1237.  Similar activity levels have been reported for the Sun-like star $\xi$ boo A ($T_{\rm eff} = 5600$ K, $P_{\rm rot} = 6.4$ days, Age: $\sim\,$0.2 Gyr, \citealt{2008ApJ...687.1264M,2012A&A...540A.138M}). As a reference value, the Solar S\,-\,index is $S_{\odot} \simeq 0.1783$, with a variation of $\sim 0.02$ from solar maximum to minimum \citep{2007ApJS..171..260L}. 

Using the mean derived values of the S\,-\,index we can now apply a transformation to obtain the parameter $R_{\rm HK}$, which takes into account the colour of the star in the activity estimation \citep{1982A&A...107...31M}. $R_{\rm HK}$ is defined as

\begin{equation}\label{eq_4}
R_{\rm HK} = (C_{\rm CF})(S_{\rm H})(1.34 \times 10^{-4})\mbox{ ,}
\end{equation}

\noindent where $C_{\rm CF}$ is a colour-dependent function. For main sequence stars with $0.3 \le (B - V) \le 1.6$, $C_{\rm CF}$ is given by

\begin{equation}\label{eq_5}
\log(C_{\rm CF}) = 0.25(B - V)^3 - 1.33(B - V)^2 + 0.43(B - V) + 0.24\mbox{ .}
\end{equation}

\noindent \cite{1984ApJ...279..763N} derived an expression in order to include photospheric corrections to the CaII core fluxes, $R^{\prime}_{\rm HK}$, which is written as  

\begin{equation}\label{eq_6}
R^{\prime}_{\rm HK} = R_{\rm HK} - R_{\rm phot}\mbox{ ,}
\end{equation}

\noindent with $R_{\rm phot}$ expressed also as a function of $(B - V)$:

\begin{equation}\label{eq_7}
\log(R_{\rm phot}) = -4.898 + 1917(B - V)^2 - 2.893(B - V)^3\mbox{ ,}
\end{equation}

\noindent and valid in the range of $ 0.44 < (B - V) \lesssim 1.0$. Table \ref{tab_4} summarises the activity indicators for the available observations of HD 1237. The errors quoted are the mean measurement errors for these quantities. $B$ and $V$ magnitudes were taken from the Hipparcos catalog (B = 7.335 mag, V = 6.578 mag, \citealt{2010MNRAS.403.1949K}). High activity levels are related with magnetic fields on the stellar surface. These magnetic signatures are encoded in the polarised spectra of the star and will be discussed in the following section.

\begin{table}[h]
\caption{Average activity indicators for HD 1237.}             
\centering                          
\begin{tabular}{c c c}        
\hline\hline                
$\left<S_{\rm H}\right>$ & $\log(R_{\rm HK})$ & $\log(R^{\prime}_{\rm HK})$ \vspace{2pt}\\    
\hline                        
$0.462 \pm 0.019$ & $- 4.29 \pm 0.04$ & $- 4.38 \pm 0.05$  \\      
\hline                                   
\end{tabular}
\label{tab_4}      
\end{table} 

\section{Magnetic field signatures}\label{sec_mag_signatures}

\begin{figure*}[!ht]
\centering 
\includegraphics[trim=4.1cm 1.4cm 2.2cm 1.4cm, clip=true, width=\hsize]{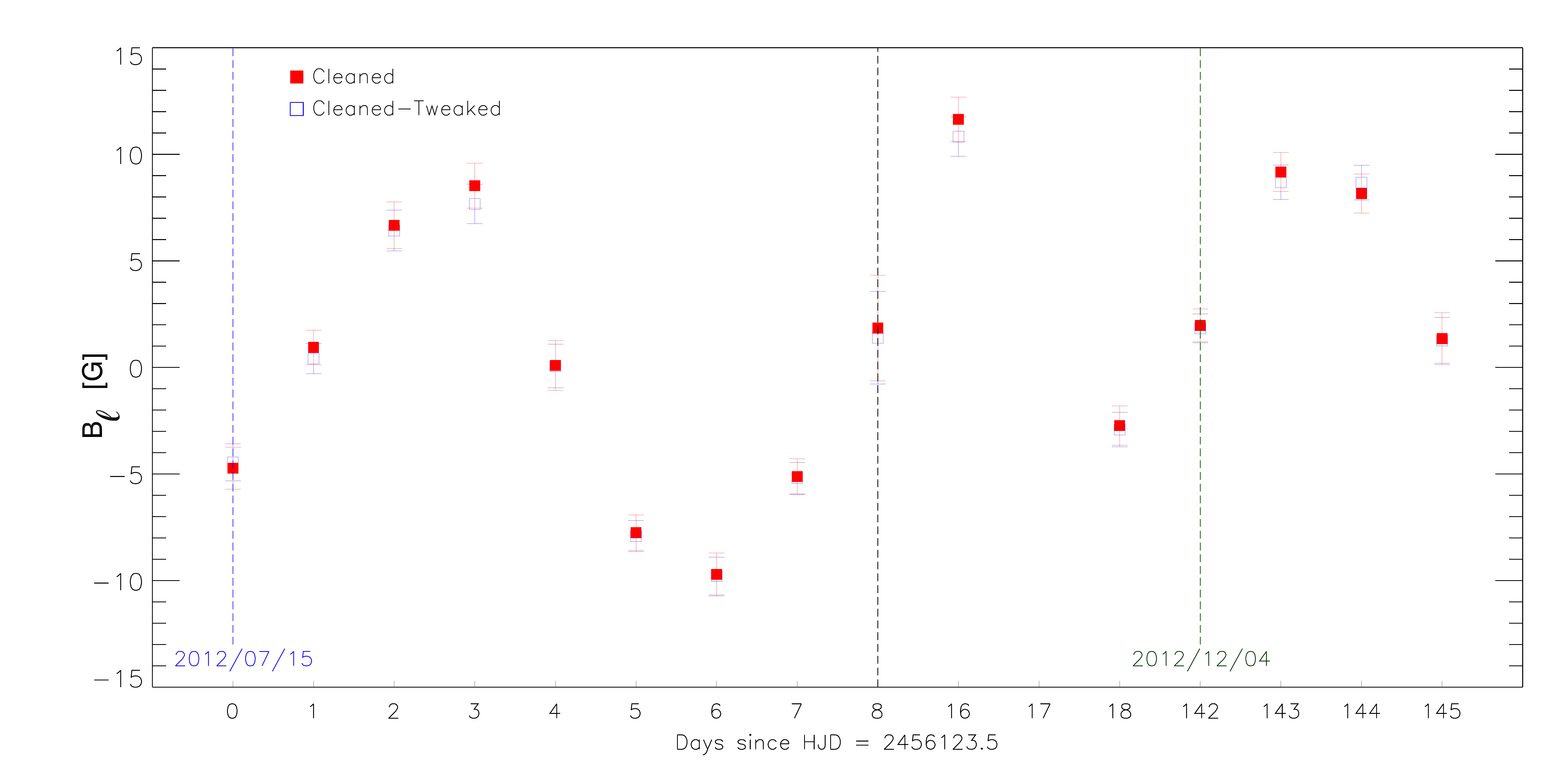}
\caption{Calculated $B_{\ell}$ for the available observations of HD 1237. The blue and green vertical lines denote the beginning of each observed epoch. The x-axis contains the number of days since the Heliocentric Julian Date (HJD = 2456123.5) of the first observation (2012 Jul 15). The last two data points of the first epoch are unevenly distributed (after the black line). Colours indicate the line mask used for the LSD procedure.}
\label{fig_6}
\end{figure*}

\noindent The signal-to-noise ratio (S/N) in the observations is not sufficiently high to detect magnetically-induced spectro-polarimetric signatures in single lines. However, by applying a multi-line technique, e.g. Least Squares Deconvolution (LSD, \citealt{1997MNRAS.291..658D}), it is possible to increase the S/N by a factor of $\sim$$50 - 100$, adding-up the signal from thousands of spectral lines over the entire spectral range (for HARPS: 378\,-\,691 nm), enhancing our sensitivity to magnetic signatures in the observations (see \cite{2010A&A...524A...5K} for a recent review of the LSD technique). This procedure requires a photospheric model (line list) matching the spectral type of our target star. This is done using an atomic line list database\footnote[2]{\url{http://vald.astro.uu.se/} -- Vienna Atomic Line Database (VALD3)} \citep{2000BaltA...9..590K} and the stellar fundamental properties listed in table \ref{tab_1}. We assumed a micro-turbulence parameter of 1.3 km s$^{-1}$ \citep{2010ApJ...720.1290G} and solar abundance for the photospheric line list which included $\sim$15000 lines within the HARPS spectral range.  

From this initial photospheric line list we generated two different masks used with the LSD calculation. In the first mask the strong lines (and lines blended with these lines) that form in the chromosphere or that break the basic assumptions of LSD, (e.g., Ca II H\&K, H$\alpha$) are removed (cleaned mask). This reduced the number of lines included to $\sim$11000. After the mask cleaning, a numerical routine based on the Levenberg-Marquardt, non-linear least-squares algorithm from the {\sc mpfit} library \citep{more78, markwardt09} is applied to fit the line mask to the observed Stokes I spectrum, adjusting the individual depths of the spectral lines (cleaned-tweaked mask). This is performed through the entire HARPS wavelength coverage. This step is more commonly carried out when applying LSD to hot (OB-type) stars which have fewer lines (e.g, \citealt{2012MNRAS.426.2738N}). Finally, LSD was applied to the spectro-polarimetric data using the final masks, generating in this way a single, averaged line profile per observation (LSD Stokes I, V and diagnostic N profiles). A velocity step $\Delta v = 1.4$ km s$^{-1}$ was used to construct the LSD profiles. This velocity spacing considers two pixels per spectral element of the instrument (in the case of HARPSpol, $R = 2.5$ km s$^{-1}$ and $3.4$ px per resolution element).

Both procedures lead us to consistent results in the obtained LSD signatures of the star. While no clear change is observed in the null polarisation check, subtle qualitative differences appear in both Stokes profiles, in the sense that the cleaned-tweaked mask seems to get a broader unpolarised profile with a slightly weaker signature in the circular polarised profile, in comparison with the clean line mask. These minimal shape differences in the LSD line profiles can have a much larger effect in hot stars where fewer lines are generally available. On the other hand, the resulting S/N of the LSD profiles for each individual observation was systematically higher using the cleaned-tweaked mask than in the cleaned case, despite the same number of spectral lines in their masks (11048). On average, a $\sim$7$\%$ increase was obtained in the S/N of the LSD profiles with the cleaned-tweaked mask. 

\subsection{Longitudinal magnetic field}\label{B_los}

\begin{figure*}[!ht]
\centering 		
\includegraphics[trim=1.44cm 0cm 0.8cm 1cm, clip=true, scale=0.51]{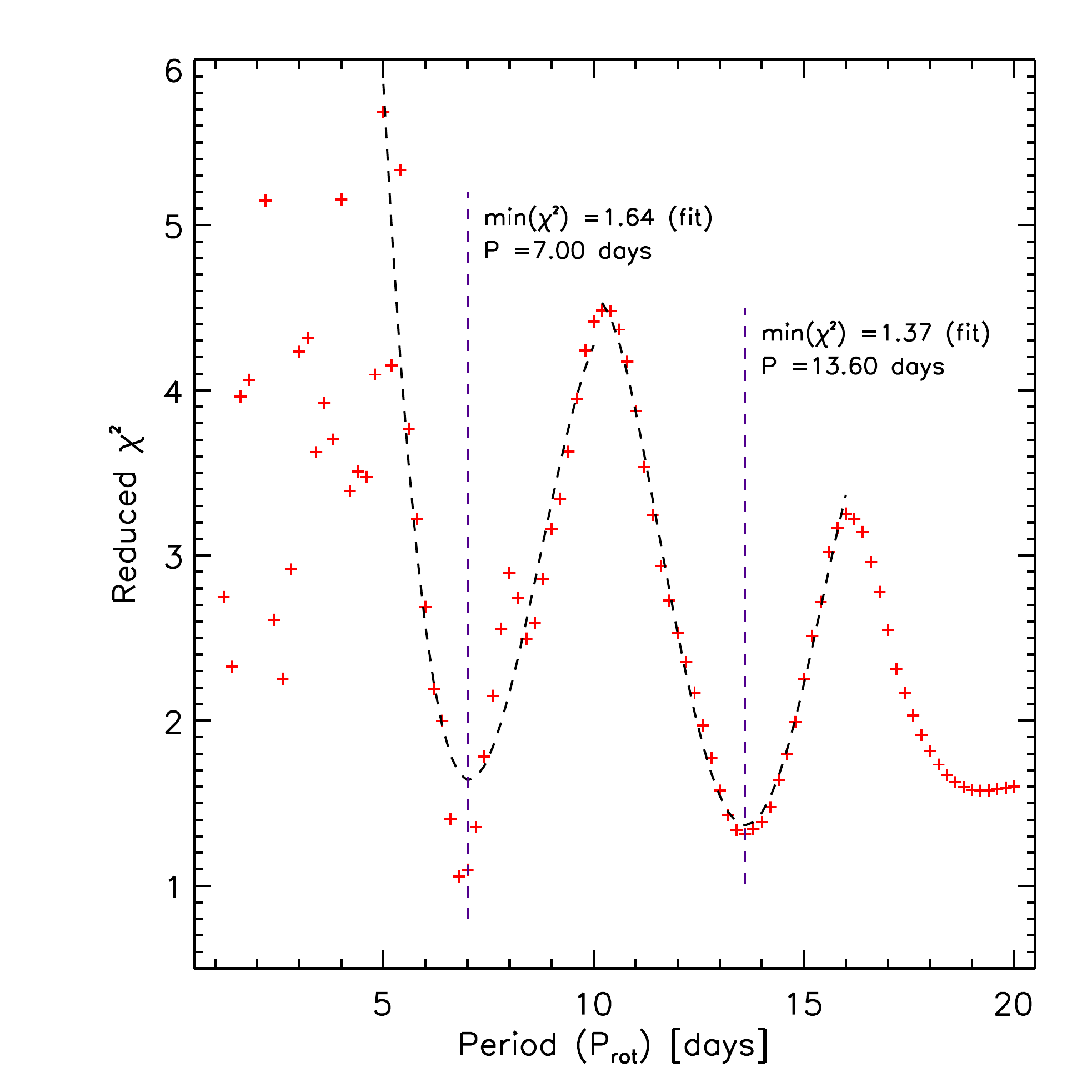} \includegraphics[trim=1.4cm 0cm 0.8cm 1cm, clip=true, scale=0.51]{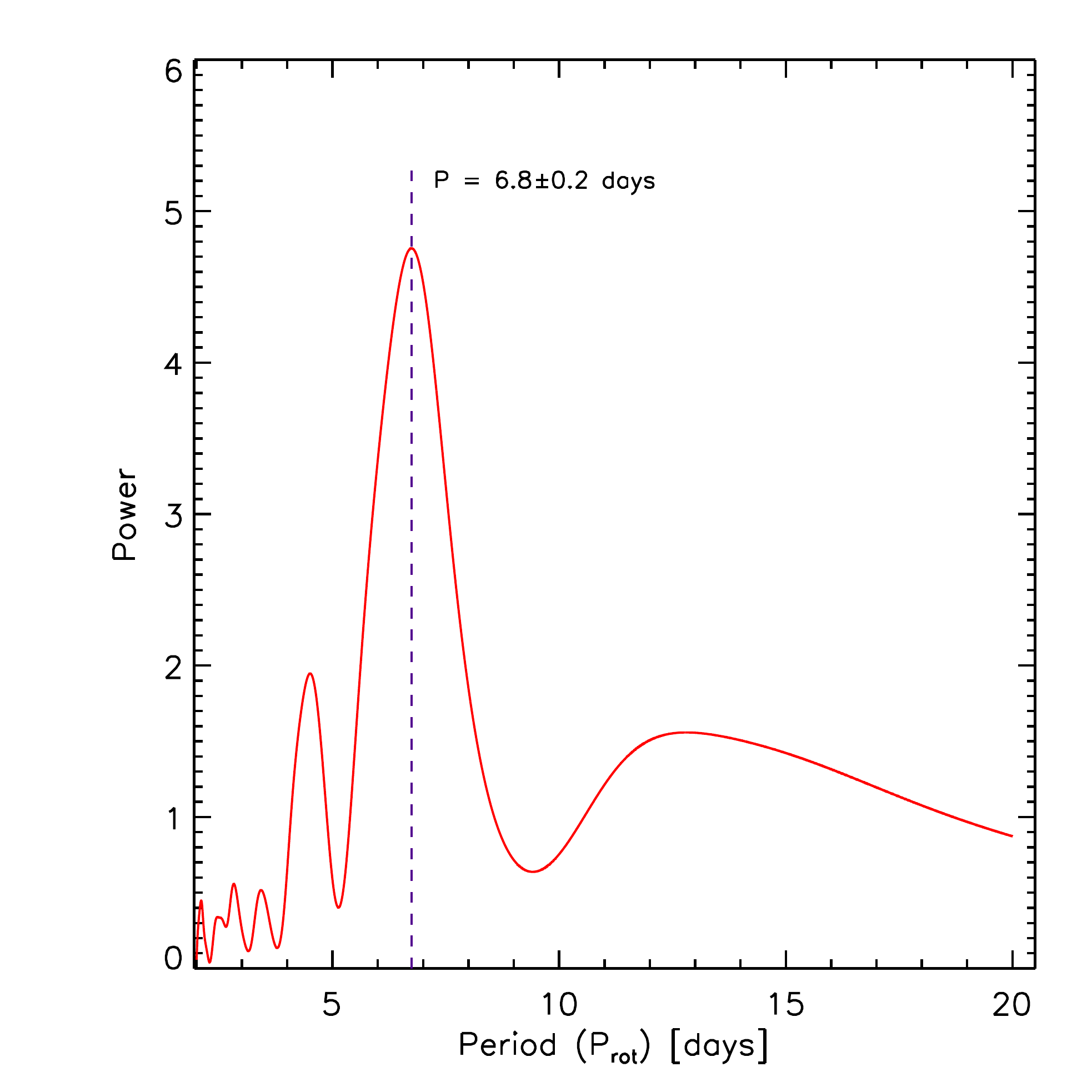}
\caption{\textit{Left:} Minimisation results for HD 1237 showing the reduced $\chi^2$ as a function of $P_{\rm rot}$. The black segmented lines corresponds to a fourth-order polynomial fit. \textit{Right:} Power spectrum obtained from the Lomb-Scargle periodogram analysis on the temporal variation of $B_\ell$ (Fig. \ref{fig_6}).}
\label{fig_7}
\end{figure*}  

\noindent Using the derived Stokes I and V LSD profiles, it is possible to obtain information about the surface averaged longitudinal magnetic field ($B_{\ell}$). Following \cite{2009ARA&A..47..333D}, an estimate of $B_{\ell}$ (in G) is given by

\begin{equation}\label{eq_9}
B_{\ell} = -714 \dfrac{\int v \mbox{V} (v) dv}{\lambda\bar{g}\int \left[1 - \mbox{I} (v)\right] dv} \mbox{ ,}
\end{equation}

\noindent where the radial velocity shift $v$ (in km s$^{-1}$) is measured with respect to the average line derived from LSD, with central wavelength $\lambda$ (in $\mu$m) and mean Land\'e factor $\bar{g}$. As this measurement is an integrated quantity over the visible surface, it cannot provide complete information for stars that host complex large-scale magnetic fields. From multiple measurements of $B_{\ell}$ taken over a stellar rotation period it is possible to gain a first insight concerning inhomogeneities of the disk-integrated magnetic field. Hence, it is also possible to estimate the stellar rotation period, using the modulation in a time-series of $B_{\ell}$ measurements, provided that they span more than one rotation period.

Figure \ref{fig_6} shows the measurements of $B_\ell$ for the available observations of HD 1237. The integration was centered in the radial velocity of the star ($- 5.2$ km s$^{-1}$), and covering the entire Stokes V signature ($\pm 12.5$ km s$^{-1}$). The uncertainties were computed from standard error propagation from the spectra. The colours denote the line mask used in the LSD procedure, showing that the slight differences in each of the measurements are consistent within the errors. This implies that either set of LSD profiles can be used to obtain robust longitudinal magnetic field measurements. As the two sets of profiles are so similar, there is only a difference in S/N without any noticeable impact on the maps structure.

$B_\ell$ shows a clearer rotational variation in comparison with $S_{\rm H}$ (Fig. \ref{fig_4}), reflecting the additive nature of the chromospheric emission, in contrast with the mixed polarity effects of the magnetic field. The behaviour of the longitudinal field is consistent through the observed epochs, with an varying amplitude of roughly $\sim$10 G. This value is somewhat larger in comparison with the solar value ($B_\ell < 4$ G, \citealt{2006AJ....131..520D,1998ApJS..116..103K}) and the average value from snapshot observations of other Sun-like stars of the same and different spectral types (3.3, 3.2 and 5.7 G for F, G and K-dwarfs respectively, \citealt{2014MNRAS.444.3517M}). However, caution is advised in these averaged comparisons given the rotational variability of $B_{\ell}$ and the nature of a snapshot survey. Similar variations have been observed in the long-term monitoring of the active Sun-like star $\xi$ Boo A \citep{2012A&A...540A.138M}.

\section{Surface magnetic field mapping}\label{sec_zdi}

\subsection{Optimal line profile and stellar parameters}\label{sec_parameters}

\noindent We reconstructed surface magnetic field maps by applying the tomographic inversion technique of Zeeman Doppler Imaging (ZDI) \citep{1987ApJ...321..496V, 1989A&A...225..456S}. ZDI, described by \cite{2000MNRAS.318..961H}, has been used to recover magnetic field maps on the surfaces of stars, ranging from T Tauri stars to binary systems (e.g. \citealt{2004MNRAS.348.1321B, 2008MNRAS.387..481D}). The code recovers the magnetic flux distribution across the stellar disk, modulated by the stellar rotation, by using time series of photospheric absorption line profiles (LSD Stokes I) and circularly polarised profiles (LSD Stokes V).  

In order to recover reliable magnetic field maps, it is necessary to properly model the LSD Stokes I and V profiles and their temporal variation. To do this, good constraints should be obtained on the local line profile description and the stellar parameters. Two different synthetic line shapes were tested: a Gaussian profile, which is commonly used in magnetic field studies in Sun-like stars (e.g. \citealt{2015A&A...573A..17B}), and a Milne-Eddington profile, fitted to a solar LSD profile derived from a high S/N HARPS spectrum. This last approach was previously considered in ZDI of accreting T Tauri stars \citep{2008MNRAS.386.1234D} and M-dwarfs \citep{2008MNRAS.390..567M}. For both cases, we assumed a linear dependance of the continuum limb darkening with the cosine of the limb angle (slope $u \simeq 0.65$, \citealt{2010A&A...510A..21S}). 

\begin{figure*}[ht]
\centering	
\includegraphics[trim=1.3cm 0.5cm 0.8cm 1cm, clip=true,scale=0.52]{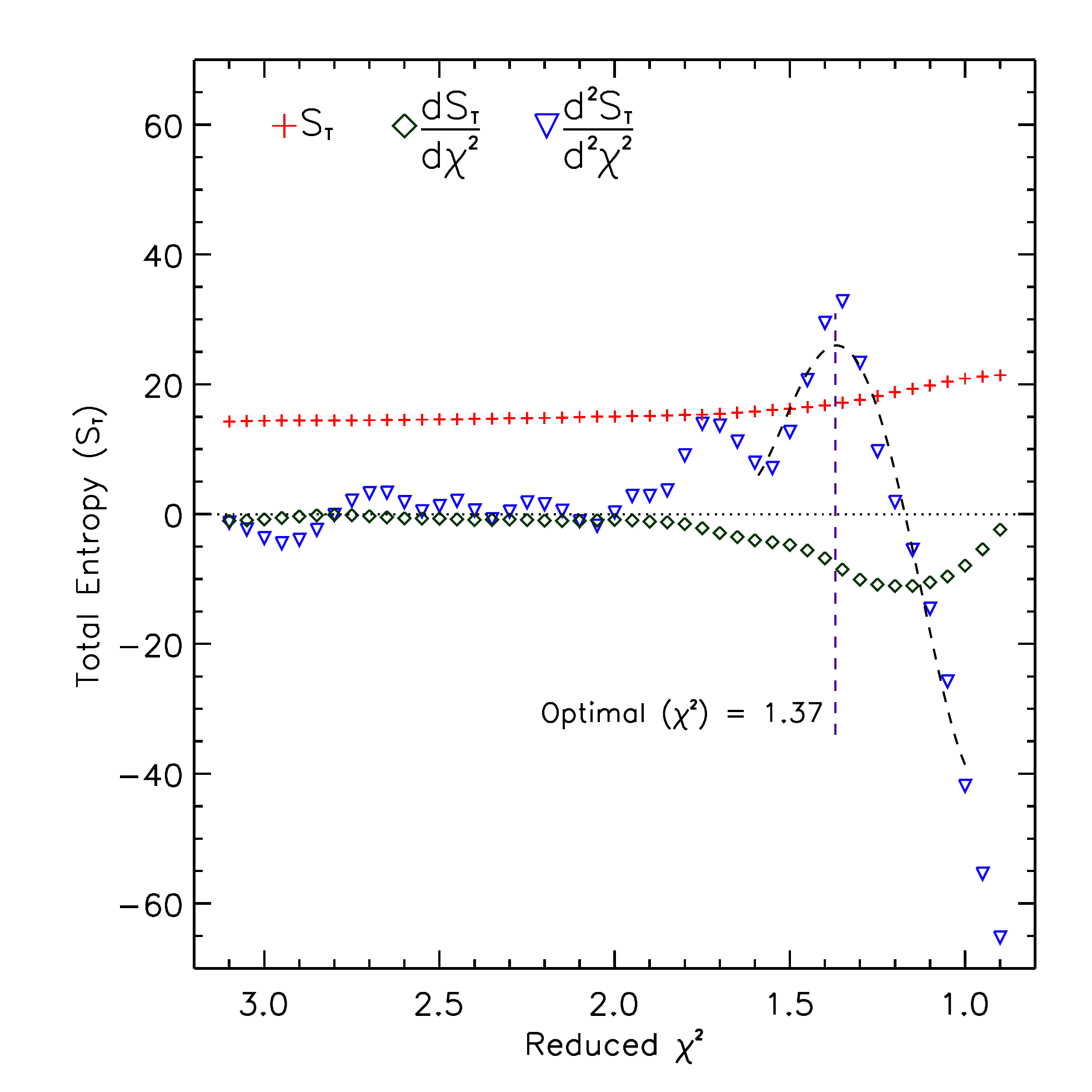} 
\includegraphics[trim=3.5cm 0.5cm 0.8cm 1cm, clip=true,scale=0.52]{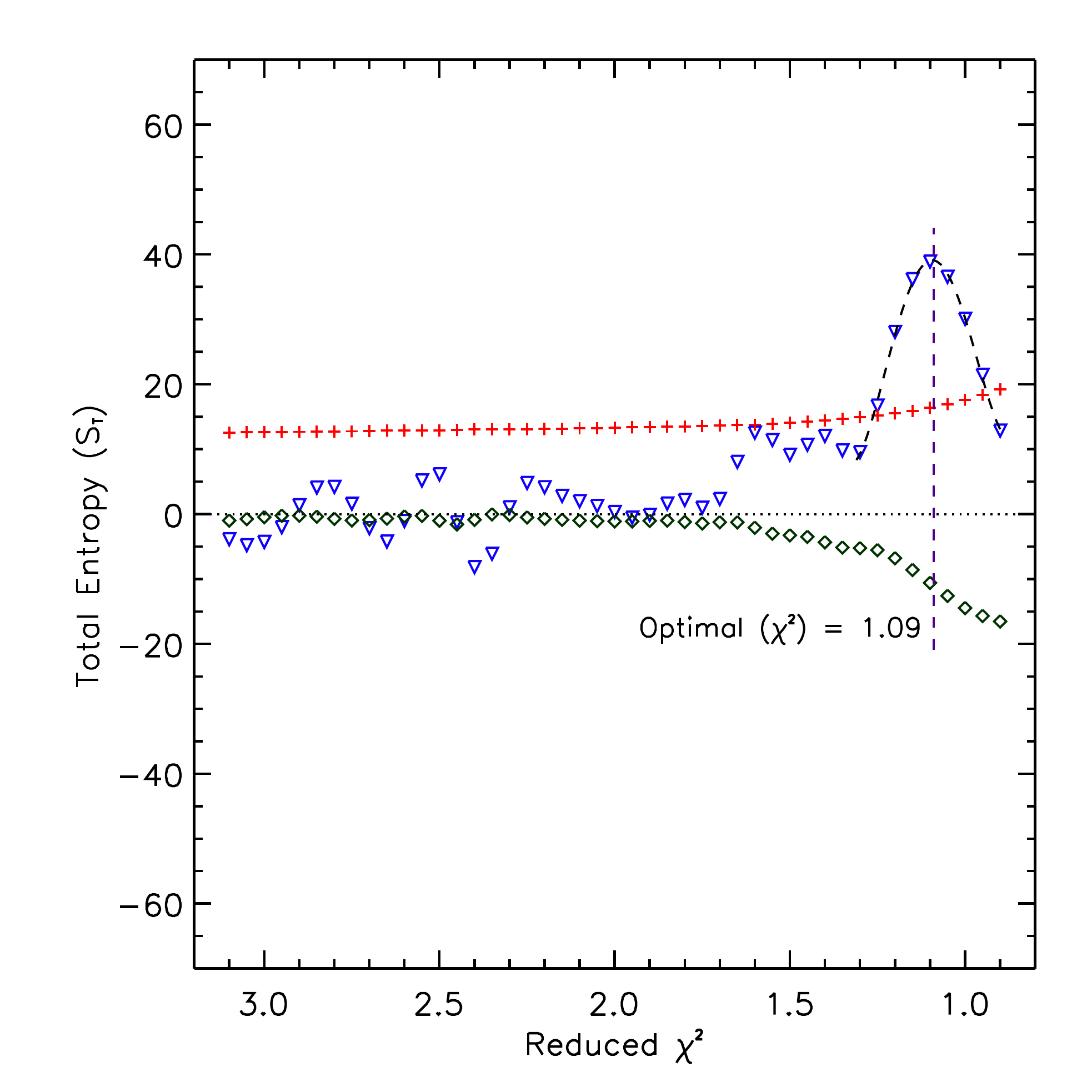} 
\caption{Optimal $\chi^2$ selection criteria applied to the ZDI solution set of HD 1237. Each panel contains the results for the July dataset using the Gaussian (left) and Milne-Eddington (right) line profiles. The red symbols show the behaviour of the total entropy content $S_{\rm T}$ as a function of the reduced $\chi^2$. Each point corresponds to a converged ZDI solution. Green and blue symbols represent the first and second derivatives as indicated. A fourth-order polynomial fit has been applied (segmented line) to find the value for which the rate of change in the information growth (second derivative) in the ZDI solution set is maximised. The optimal fit level is indicated in each case.}
\label{fig_11}
\end{figure*}

In addition, we estimated the rotational period $P_{\rm rot}$, differential rotation profile (e.g. $\Omega(l) = \Omega_{\rm eq} - d\Omega\sin^2(l)$, see \citealt{2002MNRAS.334..374P}), and inclination angle $i$ of the star. This is done by generating a grid of ZDI models, covering a range of values for the involved quantities, and minimising the reduced $\chi^2$ from synthetic line profile fitting (see \citealt{1995MNRAS.275..534C, 2009MNRAS.398..189H}). Figure \ref{fig_7} (left) shows the results of the minimisation analysis over $P_{\rm rot}$. As was mentioned in Sect. \ref{sec_properties}, the rotation period of HD 1237 is not well known. Our analysis shows the harmonic behaviour of this parameter, with a fundamental value of $\sim$ 7.0 days. We estimate a 10\% error for this period determination, given the width of the minima in the left panel of Fig. \ref{fig_6}. This estimate is consistent with the high activity level measured in this star (Sect. \ref{sec_activity}). For completeness, we performed a Lomb-Scargle periodogram, using the first epoch dataset, on the temporal variations of $B_{\ell}$ (Fig. \ref{fig_6}). Figure \ref{fig_7} (right) shows the obtained power spectrum, having a best-fit period of $P_{\rm rot} = 6.8 \pm 0.2$ days with an associated P-Value statistic of $5.18 \times 10^{-5}$ \citep{2009A&A...496..577Z}.  

No good constraints were obtained by $\chi^2$ minimisations for the differential rotation parameters and the inclination angle of the star. Therefore, no differential rotation profile was included in the mapping procedure. For the inclination angle, we considered the expected value from rigid body rotation, i.e. $\sin i = (P_{\rm rot} \cdot v\sin i)/(2\pi R_*)$. Given the uncertainties of the involved quantities (see Table \ref{tab_1}), the inclination angle should lie somewhere between $\sim$ 40 - 60$\,^\circ$. We performed reconstructions for 40, 50 and 60$\,^\circ$ and find no substantial differences between these magnetic field reconstructions. We present here the maps obtained assuming a 50$\,^\circ$ inclination angle.
 
Finally, we compared the optimum stellar and line parameters associated with the LSD profiles produced using both masks (cleaned mask and clean-tweaked mask). No significant difference is found in the optimal parameters. Hence in the subsequent analysis we only consider the LSD profiles produced using the cleaned-tweaked mask. 

\subsection{Optimal Fit Quality: Entropy Content in ZDI Maps}\label{sec_criteria}

\noindent The last of part of the analysis corresponds to the selection of the optimal fit quality (reduced $\chi^2$), of the model with respect to the observations. In principle, the recovered field distribution and associated profiles should try to achieve the lowest possible value of $\chi^2$. Still, given the limitations of the observations and the ZDI technique (S/N, spatial resolution, phase-coverage, etc.), the goodness-of-fit level has to be determined carefully to avoid the appearance of numerical artifacts in the final maps. This is particularly important in the case of resolution-limited maps (i.e. slowly rotating stars). However, there are few published procedures to establish a robust ``stopping criterion'', i.e. the point at which noise starts to affect the reconstructed image. Motivated by this, we propose a systematic method for estimating the optimal fit quality for a given set of ZDI magnetic field maps, using HD 1237 as a \textit{test-case}. It is important to note here that since this procedure is defined \textit{a posteriori} over the resulting maps themselves (2D images), it does not modify the regularisation functions imposed to ZDI. In this sense, its application to other stellar systems should be straightforward. 

We begin with the definition of the \textit{entropy} $S$, as an estimate of the information content in an image\footnote[2]{A similar implementation of entropy is commonly used as a regularisation function in ZDI (see \citealt{2002A&A...381..736P}).}. Following \cite{Sonka:2007:IPA:1210103}, let $P(k)$ be the probability that the difference between two adjacent pixels is equal to $k$. The image entropy can be estimated as 

\begin{equation}\label{eq_11}
S = - \sum_k P_{k}\log_2 (P_k)\mbox{,}
\end{equation}

\noindent where $\log_2$ is the base 2 logarithm. This implies that a larger or smaller amount of entropy in the image will depend on the contrast between adjacent pixels. An image that is perfectly constant will have an entropy of zero. For the methodology described below, we are not interested in the absolute values of the entropy, rather its overall behaviour as a function of the reduced $\chi^2$.

For a given converged ZDI solution (i.e. a particular value of reduced $\chi^2$), we can calculate the total entropy content ($S_{\rm T}$) by applying the definition given by equation (\ref{eq_11}) to each one of the recovered maps:

\begin{equation}\label{eq_12}
S_{\rm T} = S_{\rm R} + S_{\rm M} + S_{\rm A}\mbox{,}
\end{equation}

\noindent where $S_{\rm R}$, $S_{\rm M}$, and $S_{\rm A}$ represent the entropy contained in the radial, meridional, and azimuthal field components maps, respectively. Figure \ref{fig_11} shows in red the behaviour of $S_{\rm T}$ as a function of the reduced $\chi^2$ for the ZDI solution set of HD 1237. Each point in Fig. \ref{fig_11} corresponds to a converged ZDI solution.

As expected, by decreasing the reduced $\chi^2$ the information content in the resulting ZDI solution increases (field strength and structure). The green and blue symbols indicate the entropy growth (first derivative) and its variation (second derivative), respectively. The overall behaviour of the total entropy content and information growth is consistent for both cases. In the case of the Gaussian profile (Fig. \ref{fig_11}, left panel), the entropy growth remains fairly constant (close to zero) for large reduced $\chi^2$ values, reaching a maximum\footnote[2]{The apparent negative sign in the first derivative is due to the reversed direction of the x-axis (reduced $\chi^2$).} around $\chi^2 \simeq 1.2$. However, by that point the concavity of the curve has changed (negative second derivative), suggesting a different regime for the information growth in the ZDI solutions. This is interpreted as a noise signature, reflected as artifacts in the final maps leading to an additional increment in the information growth. For this reason we adopt as optimal fit level, the reduced $\chi^2$ value for which the rate of change in the information growth (second derivative) in the ZDI solution set is maximised. In this particular case, this occurs around $\chi^2 \simeq 1.4$, as is indicated by the fourth-order polynomial fit (segmented line) in the left panel of Fig. \ref{fig_11}. A similar criterion plot is constructed for the Milne-Eddington line profile (Fig. \ref{fig_11}, right panel), where a lower optimal reduced $\chi^2 \simeq 1.1$ is achieved in this case\footnote[3]{The same criterion was applied to generate the magnetic field maps in the December dataset (Appendix \ref{app_1}). An optimal reduced $\chi^2 \simeq 0.6$ was obtained in this case, as is expected for a dataset with fewer constraints.}. 

\subsection{ZDI Maps and Synthetic Stokes V Profiles}\label{sec_ZDI_maps}

\noindent We reconstruct the ZDI surface magnetic field maps and the synthetic circularly polarised profiles based on the time-series of LSD Stokes V spectra. For this we use the cleaned-tweaked line mask (Sect. \ref{sec_mag_signatures}), and the stellar and line parameters derived in Sect. \ref{sec_parameters}. The goodness-of-fit level in each case is selected under the criterion described in the last section. 

\begin{figure}[!ht]
\centering		
\includegraphics[trim=2cm 2.1cm 3.0cm 0cm, clip=true, width=\hsize]{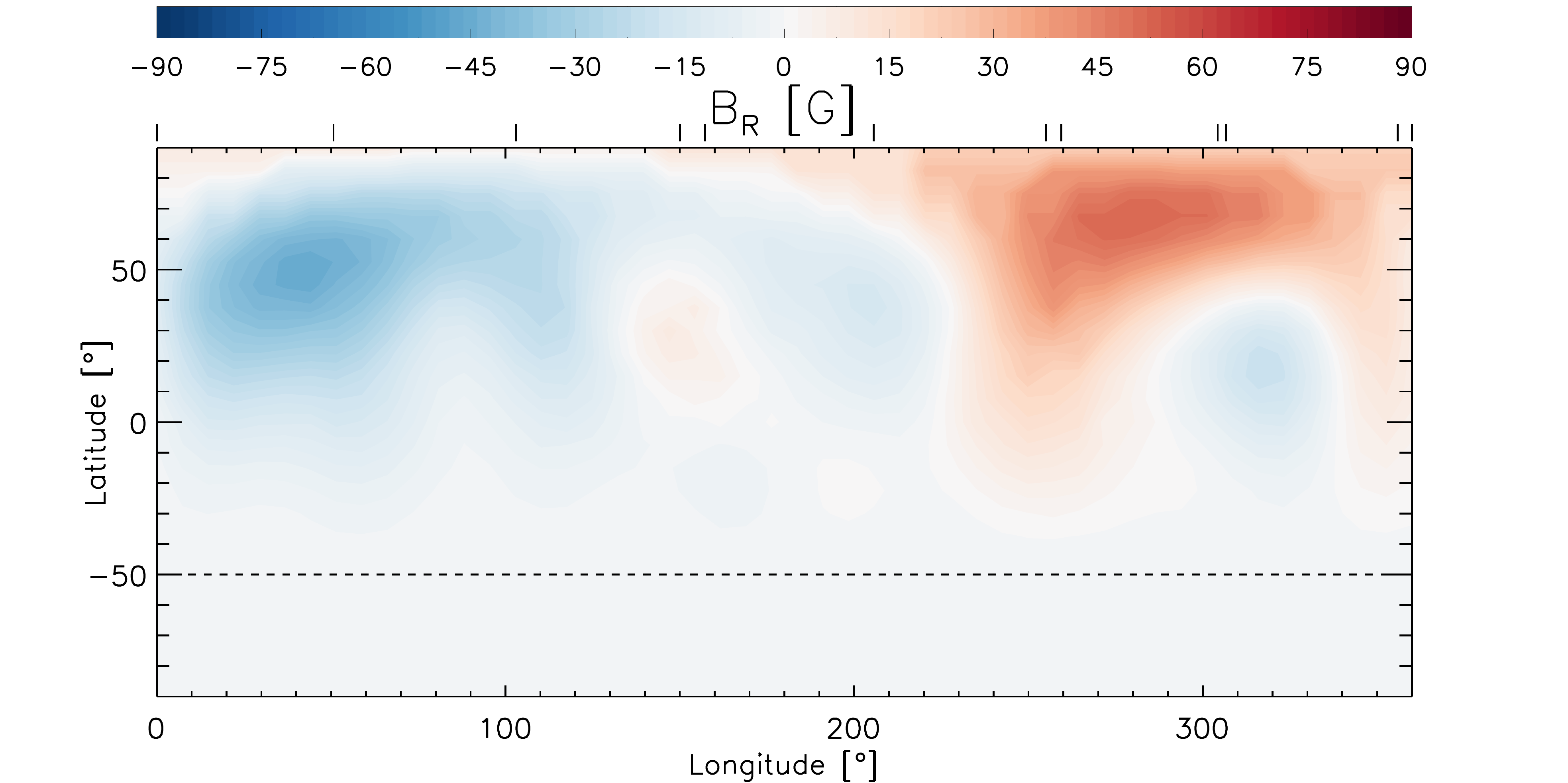} 
\includegraphics[trim=2cm 2.1cm 3.0cm 2cm, clip=true, width=\hsize]{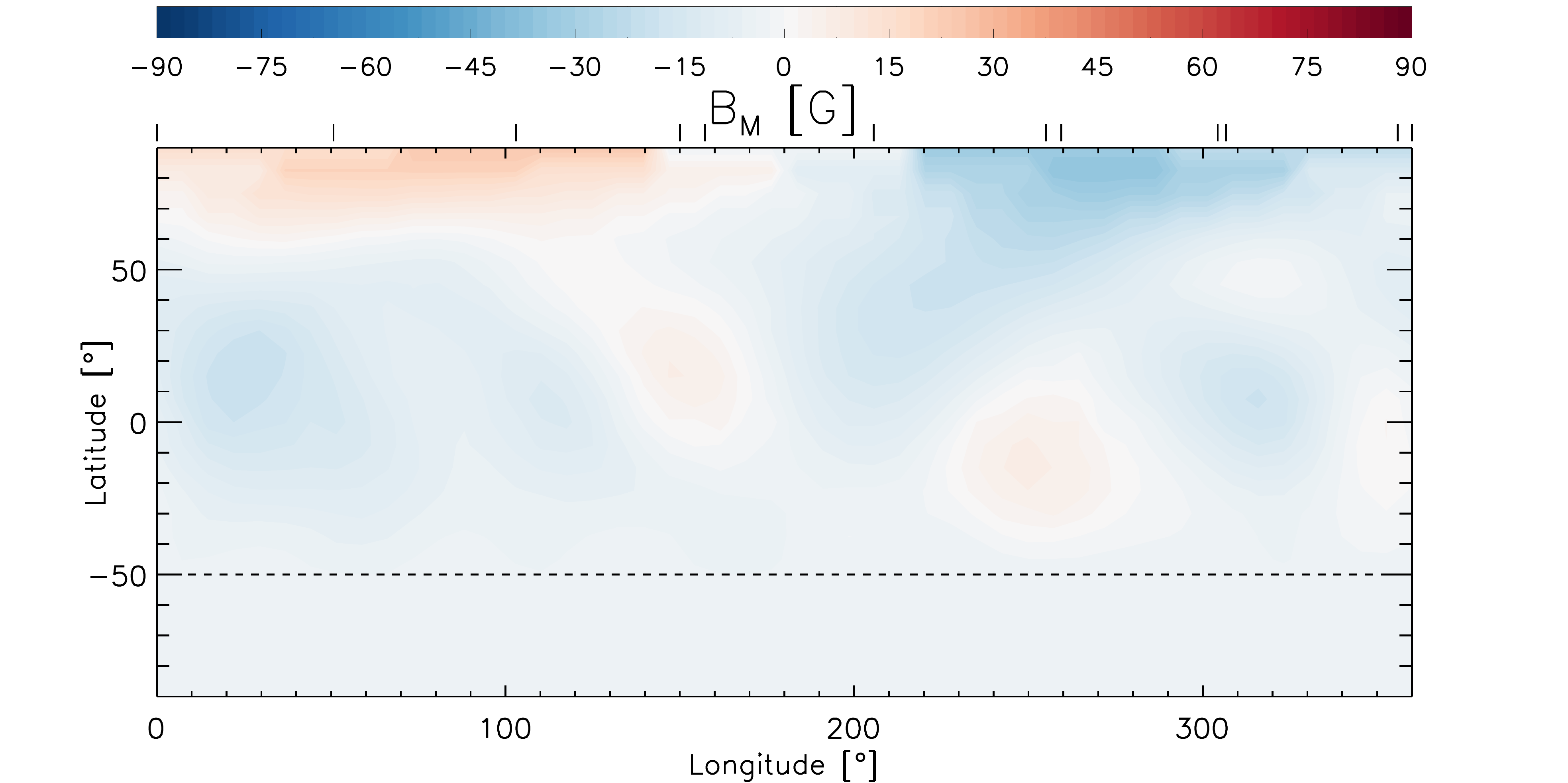} 
\includegraphics[trim=2cm 0cm 3.0cm 2cm, clip=true, width=\hsize]{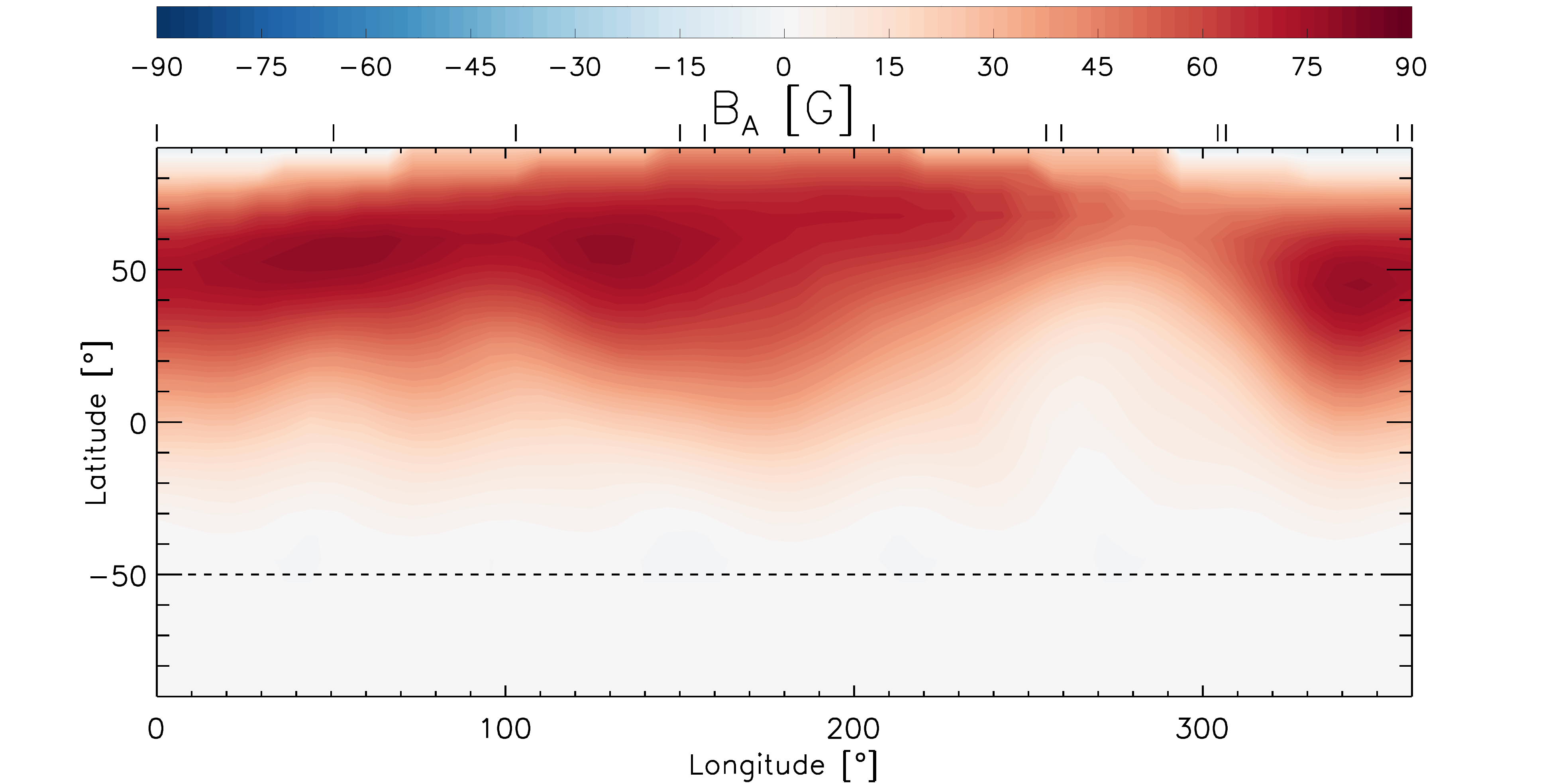} 
\includegraphics[trim=2.0cm 2cm 1.0cm 0.2cm, clip=true, width=\hsize]{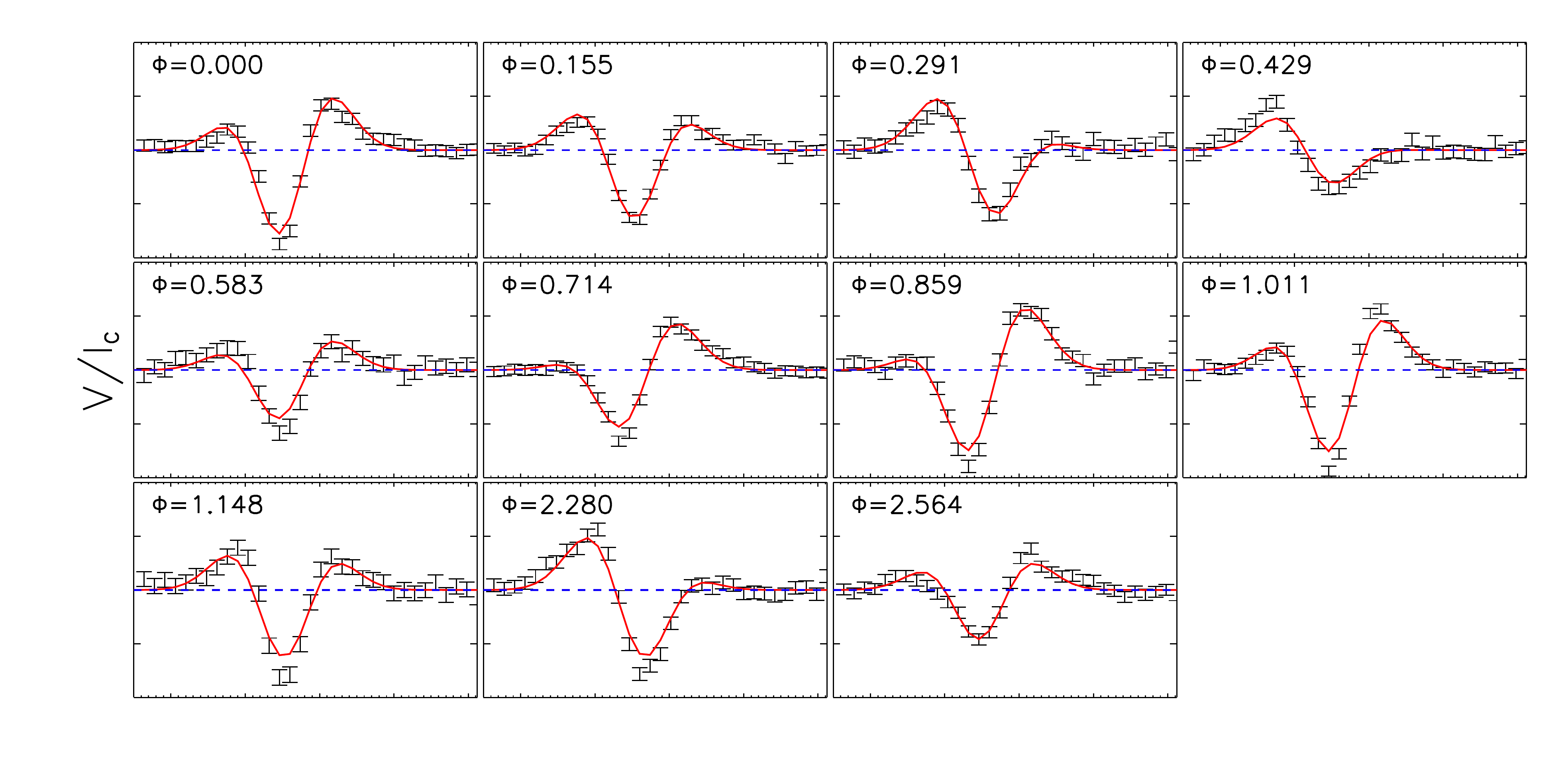} 
\caption{Results of the ZDI analysis for the first epoch observations of HD 1237 using the Gaussian line profile. The first three panels show the surface magnetic field components B$_{\rm R}$, B$_{\rm M}$, and B$_{\rm A}$, respectively. The colour scale indicates the polarity and the magnitude of the magnetic field component in G, while the phase coverage is indicated by the black tick marks in the upper y-axis. The segmented horizontal line indicates the surface visibility limit, imposed by the adopted inclination angle of the star ($i = 50\,^{\circ}$). The last panel show the comparison between synthetic (red) and observed (black) Stokes V profiles obtained for this particular epoch, in each observational phase $\Phi$, where the recovered maps fit the spectro-polarimetric data to an optimal reduced $\chi^2 = 1.4$.}
\label{fig_8}
\end{figure}

\begin{figure}[!h]
\centering		
\includegraphics[trim=2cm 2.1cm 3.0cm 0cm, clip=true, width=\hsize]{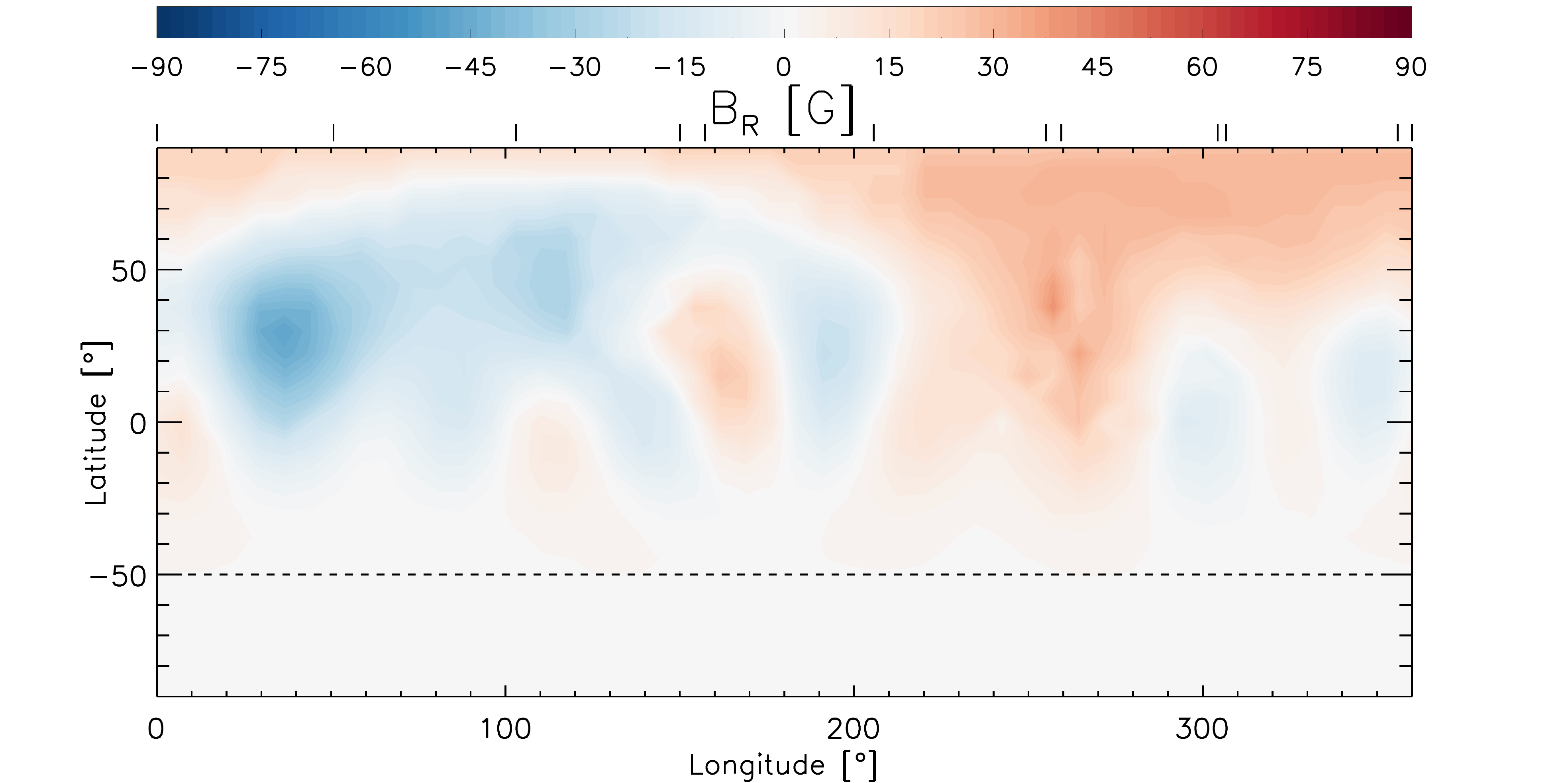} 
\includegraphics[trim=2cm 2.1cm 3.0cm 2cm, clip=true, width=\hsize]{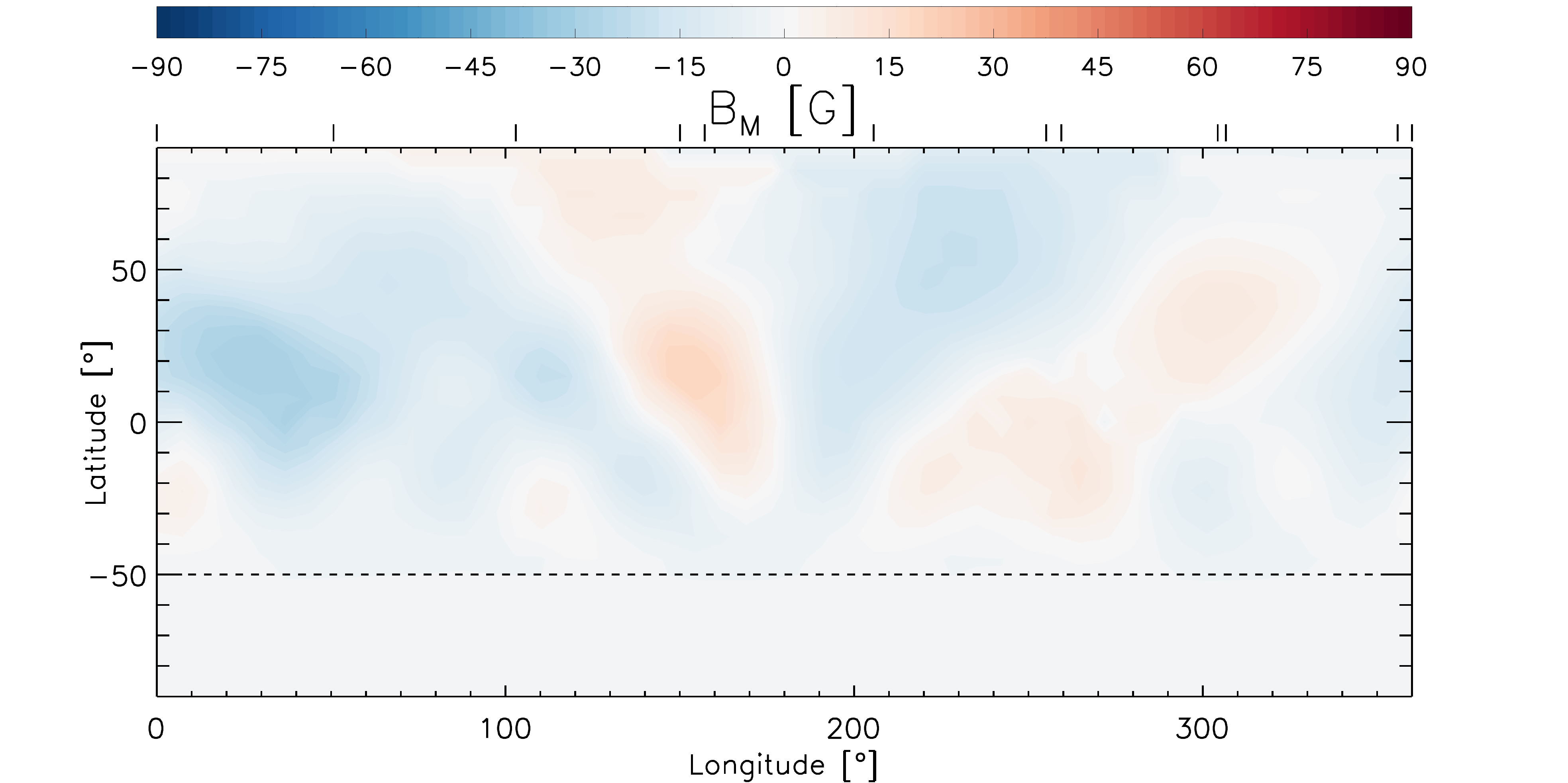} 
\includegraphics[trim=2cm 0cm 3.0cm 2cm, clip=true, width=\hsize]{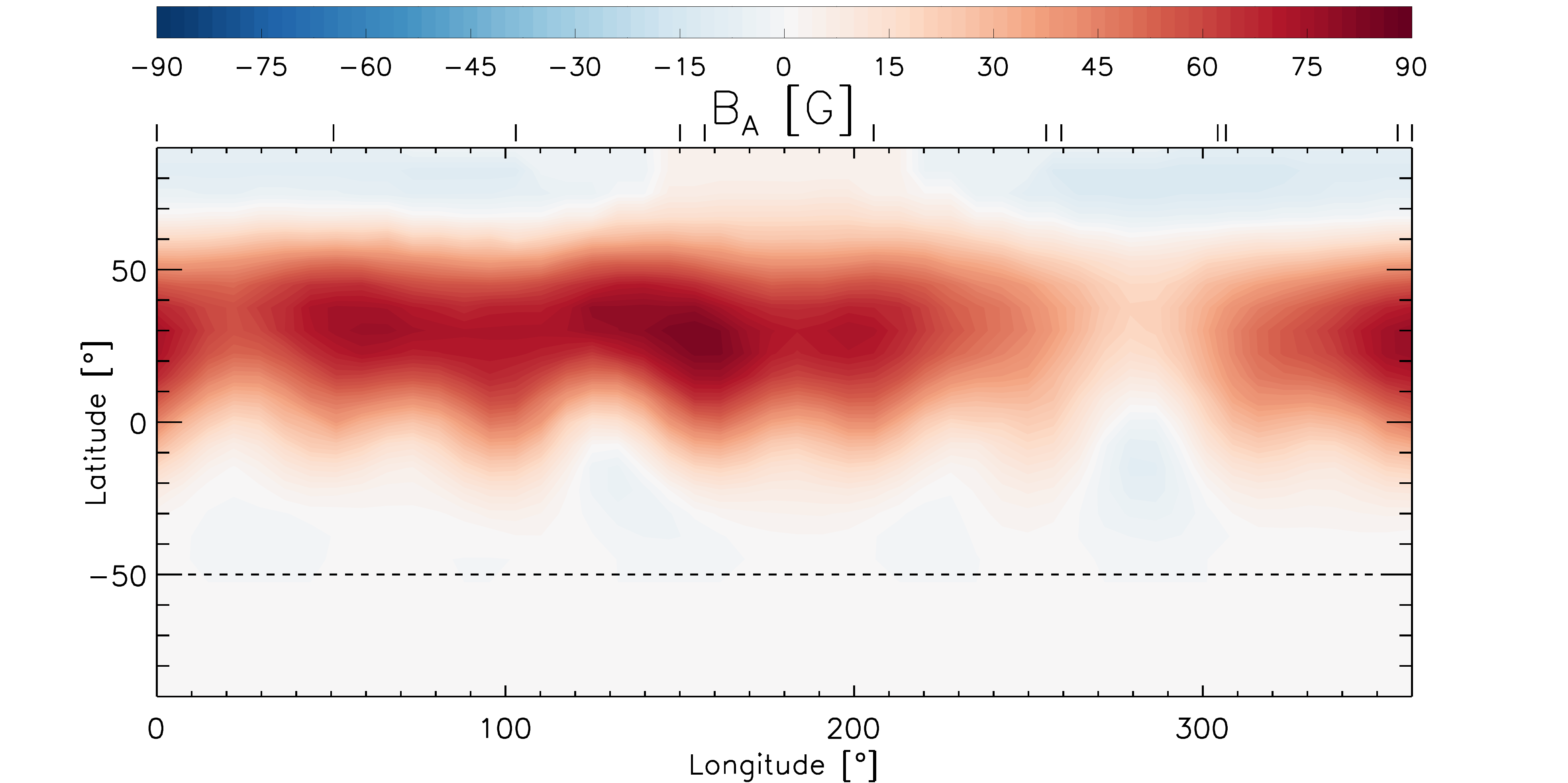} 
\includegraphics[trim=2.0cm 2cm 1.0cm 0.2cm, clip=true, width=\hsize]{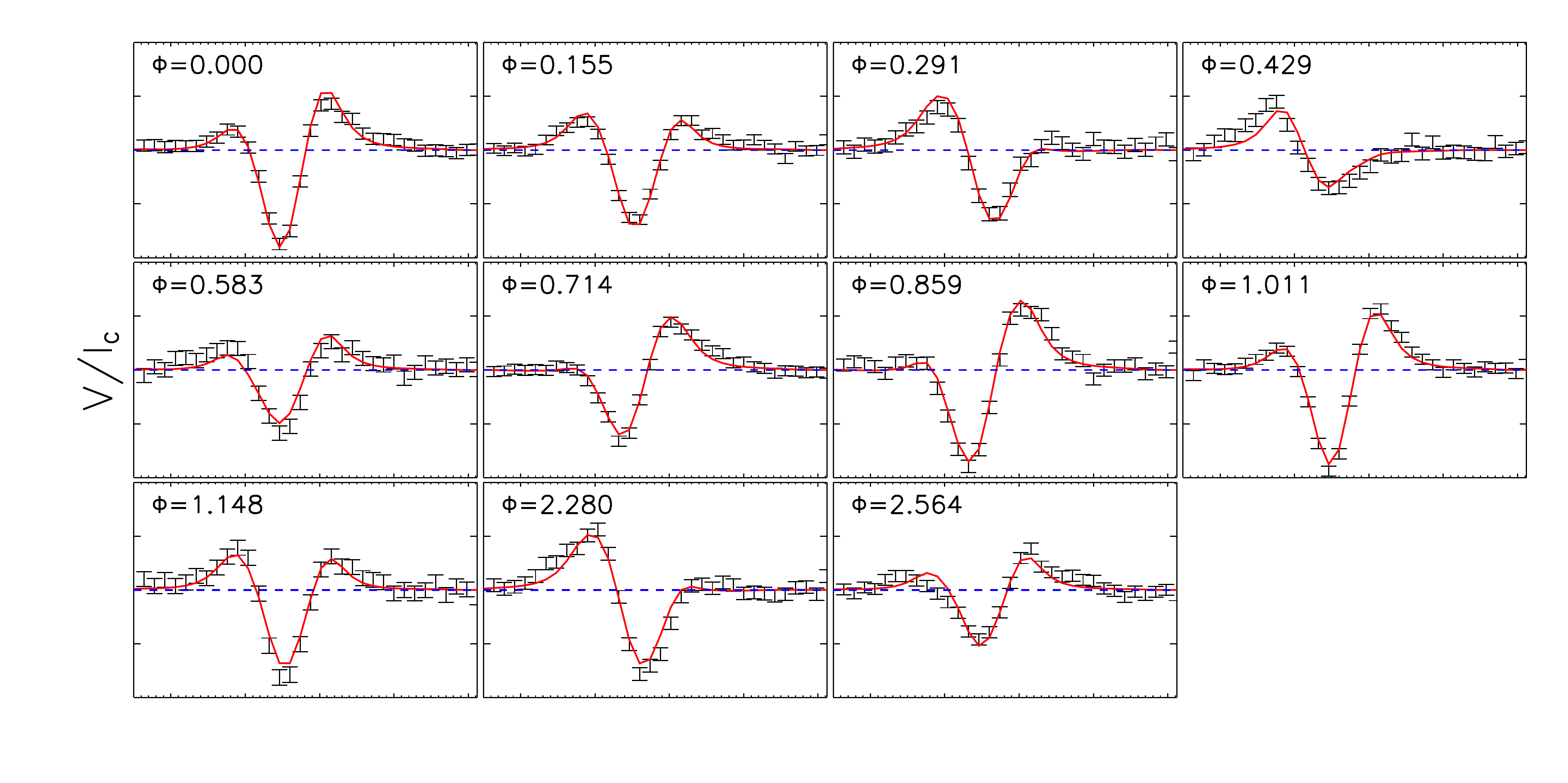} 
\caption{Results of the ZDI analysis for the first epoch observations of HD 1237 using the Milne-Eddington line profile. See caption of Figure \ref{fig_8} for more details. In this case, the maps fit the spectro-polarimetric data to an optimal reduced $\chi^2 = 1.1$.}
\label{fig_9}
\end{figure}

Figures \ref{fig_8} and \ref{fig_9} show the results of the ZDI procedure for the first epoch observations (2012 July) of HD 1237. Four vertical panels are presented, where the first three correspond to the Mercator-projected magnetic field maps in Gauss (G), decomposed into the radial (B$_{\rm R}$ - top), meridional (B$_{\rm M}$ - middle), and azimuthal (B$_{\rm A}$ - bottom) components. The phase coverage is indicated by the black tick marks in the upper y-axis. The spatial resolution of the maps is $\sim18^{{\circ}}$ in longitude at the stellar equator (at the poles, the map has a considerably poorer resolution than the equator). However, the phase coverage also has a significant impact on the level of detail that can be recovered. In this case, the right hand side would have a slightly better resolution than the left hand side of the image. 
The last panel shows the fitted synthetic Stokes V profiles to the spectro-polarimetric observations in each rotational phase ($\Phi$). 

The recovered maps show a relatively complex field distribution across the surface. The field is dominated by the azimuthal component, displaying a strong ($\sim +90$ G) large-scale ring-like structure around 45\,$^\circ$ in latitude. In the radial component, two large and moderately strong ($\sim \pm 50$ G) regions of opposite polarities are also located at higher latitudes while weaker ($\sim \pm 25$ G) small-scale features of mixed polarities appear close to the equator. These large magnetic features are somewhat preserved, with reversed polarities, in the meridional magnetic field maps. However, some cross-talk from the radial component may be present in the meridional map (see \citealt{1997A&A...326.1135D}).

Although the overall large-scale structure and field strength are consistent between both line profiles, several differences are clearly visible in the characteristics of the field components. First, the maps recovered using the Milne-Eddington line shape include additional small-scale features leading to a more complex field distribution in the surface. These are more prominent in the maps for the radial and meridional components. Second, the recovered magnetic field distribution, in the Milne-Eddington case, seems to be slightly shifted to lower latitudes. These differences can be understood from the fact that the Milne-Eddington line shape is a better representation of the derived LSD Stokes I profile. This is also true for the the shape of the circularly polarised profile, leading to a more detailed (additional small-scale structures) and a field distribution shifted to lower latitudes (as a consequence of the sensitivity in the core of the profile). Both elements are translated into the lower optimal reduced $\chi^2$ value that can be reliably achieved in this case (see Sect. \ref{sec_criteria}), and therefore an improved fit (lower panel in Fig. \ref{fig_9}).  

The reconstructed map for the second epoch observations (2012 December) and the corresponding synthetic Stokes V profiles are presented in the Appendix \ref{app_1}.

\section{Summary and discussion}\label{sec_summary}

\noindent In this paper, we presented a detailed study, covering two observational epochs, of the activity and magnetic field structure of the Sun-like planet-hosting star HD 1237. The chromospheric activity level of HD 1237 estimated from the calibrated $S_{\rm H} = 0.46 \pm 0.02$ and $\log(R^\prime_{HK}) = - 4.38 \pm 0.05$ values is similar to other active Sun-like stars (e.g. $\xi$ boo A, \citealt{2012A&A...540A.138M}), and considerably higher than the Solar case ($\sim 3$ times higher in terms of the average S-index, \citealt{2007ApJS..171..260L}). A much larger difference has been reported for the X-ray activity level, with a $\log(L_{\rm X}) = 29.02 \pm 0.06$ for HD 1237 \citep{2008ApJ...687.1339K}, which is two orders of magnitude higher than the value estimated for the Sun during solar maximum (\citealt{2000ApJ...528..537P}). The chromospheric activity level remained fairly constant over the $\sim$\,5 month period between the two sets of observations. Our estimate falls between the previous measurements of $\log(R^\prime_{HK}) = - 4.27$ \citep{2001A&A...375..205N} and $\log(R^\prime_{HK}) = - 4.44$ \citep{2005A&A...443..609S}. Given the large uncertainties in the activity indicators, it is difficult to address if these variations have some correspondence to a magnetic cycle in the star. 

In order to extract magnetic field signatures from spectro-polarimetric data, we applied the LSD multi-line technique to the observations. Two different line-masks for the LSD profiles were compared and tested. The standard procedure involves employing a mask ``cleaned'' of chromospheric and strong NLTE line profiles (e.g. \citealt{2014MNRAS.444.3517M}); we compared the results from this procedure with improving the line-list further by ``tweaking'' line-strengths, so that they are tailored to the line depths in the observed spectrum. Both approaches lead to similar results in the obtained LSD profiles, and therefore in the physical quantities inferred (e.g. the longitudinal magnetic field, $B_{\ell}$). However, for the same number of spectral lines, the average S/N of the LSD profiles recovered with the aid of the cleaned-tweaked line mask was $\sim$\,7\% higher than in the cleaned case. 

The longitudinal magnetic field ($B_{\ell}$), estimated from the Stokes I and V LSD profiles, showed a clear rotational modulation with an amplitude of $\sim$10 G. This behaviour was preserved in both observed epochs. Placing these measurements in context of other stars, similar variations have been observed in the long-term monitoring of the active Sun-like star $\xi$ Boo A ($\sim$ 4-9 G, Age: $\sim\,$0.2 Gyr, \citealt{2008ApJ...687.1264M,2012A&A...540A.138M}), the K-dwarf exoplanet host $\epsilon$ Eri ($\sim$ 10-12 G, Age: $\sim\,$0.2 - 0.8 Gyr, \citealt{2008A&A...488..771J,2014A&A...569A..79J}), and more recently for the young solar analogue HN Peg ($\sim$ 14 G, Age: $\sim\,$0.2 Gyr, \citealt{2013A&A...556A..53E, 2015A&A...573A..17B}). The chromospheric and X-ray activity levels of these stars are also very similar in comparison to HD 1237. $\xi$ Boo A has an average S-index of $S_{\rm HK} \simeq 0.45$ and $\log(L_{\rm X}) \simeq 28.91$ \citep{1996ApJ...465..945G, 2010ApJ...717.1279W}. $\epsilon$ Eri has strong magnetic activity with a mean $S_{\rm HK} \simeq 0.50$ and $\log(L_{\rm X}) \simeq 28.22$ \citep{2014A&A...569A..79J, 2011A&A...529C...1P}. Similarly, previous reports for HN Peg show a $S_{\rm HK} \simeq 0.35$ and $\log(L_{\rm X}) \simeq 29.19$ \citep{2015A&A...573A..17B, 2004A&A...417..651S}. Despite the various similarities among these systems, the complete relation between the magnetic field and its influence over different layers of the stellar atmosphere (activity) is not yet fully understood. Similar to these other systems, the chromospheric activity of HD 1237 does not show a clear correlation with $B_\ell$. This is interpreted as the result of probing different spatial and temporal scales in each of these measured quantities.  

We recovered the optimal stellar parameters of HD 1237 using a reduced $\chi^2$ minimisation scheme, based on the tomographic inversion technique Zeeman Doppler Imaging (ZDI). The analysis yields a rotation period, $P_{\rm rot} = 7.0 \pm 0.7$ days. Literature values of $P_{\rm rot}$ are uncertain, ranging between $\sim 4.1 -  12.6$ days (see \citealt{2010MNRAS.408.1606W} and references therein). The 7.0 days value found in this work is consistent with the ZDI analysis and both, the chromospheric and coronal activity levels of the star. 
Similar procedures were applied to estimate the inclination angle and the differential rotation of HD 1237. However, the available observations did not provide enough constraints for a robust determination of these parameters. Therefore, we considered an inclination angle derived using the stellar properties of the star and solid body rotation (i.e. $i\sim\,$50\,$^{\circ}$). No differential rotation profile was included in the reconstruction of the magnetic field maps.

For the surface magnetic field mapping procedure, two different synthetic line shapes (Gaussian/Milne-Eddington) were tested in order to investigate their impact on the ZDI maps for slowly rotating solar-type stars. We showed that, both profiles recover robust magnetic field maps. However, the Milne-Eddington line profile yields a better spectro-polarimetric fit (lower optimal reduced $\chi^2$) leading to a more detailed structure recovered in the ZDI maps compared to the Gaussian case. In connection to this, a fit-stopping criterion based on the information content (entropy) of the ZDI maps solution set was introduced. This allows the identification of the optimal reduced $\chi^2$ value, avoiding to some extent possible the appearance of numerical artifacts in the ZDI maps. The optimal fit level is given by the reduced $\chi^2$ value for which the rate of change of the entropy growth in the ZDI solution set is maximised. 

The large-scale magnetic field of HD 1237 showed a complex distribution at the stellar surface. The strongest magnetic field features appear at mid latitudes ($\sim\,$45\,$^{\circ}$) in the azimuthal and radial components. The field is dominated by the azimuthal component, displaying latitudinal belts or ring-like structures across the stellar surface. This has been observed in other studies of active Sun-like stars (e.g. \citealt{2013arXiv1310.2073F}) and in numerical simulations of a rapidly rotating Sun \citep{2010ApJ...711..424B}. As has been suggested previously, the appearance of significant, or even dominant toroidal fields in the surface of these types of stars is connected with the rotation period, with $\sim 12$ days as the rotation threshold \citep{2008MNRAS.388...80P}. In comparison with other cool stars, HD 1237 ($P_{\rm rot} = 7.0$ days, this work, $M_{*} = 1.0 \pm 0.1$ M$_{\odot}$, \citealt{2010ApJ...720.1290G}) would be located near $\xi$ Boo A (G8V, Age: $\sim\,$0.2 Gyr) and GJ 182 (M0V, Age: $0.50^{+1.0}_{-0.3}$ Gyr, \citealt{2004ApJ...608..526L}) in the mass-period, large-scale magnetic field diagram (Fig. 3 in \citealt{2009ARA&A..47..333D}), with a mostly toroidal field topology and a considerable contribution from the radial component. With the same spectral type as HD 1237, $\xi$ Boo A displays a slightly weaker field in its surface ($\sim\pm$60 G), with an alternating dominance between the radial and the azimuthal components in the observed long-trend evolution ($\sim$ 4 years, \citealt{2012A&A...540A.138M}). During the observed azimuthal-dominated epochs of this star, the field distribution is highly similar to the one derived for HD 1237 in this work. A strong uni-directional azimuthal field appears at low latitudes, with large mixed polarity regions in the radial field and a minor contribution from the meridional component. However, the magnetic regions of $\xi$ Boo A are much larger and less concentrated as the ones found for HD 1237. This could be related with the small difference in their rotation periods (6.43 days for $\xi$ Boo A and 7.0 days for HD 1237) and/or with the differential rotation that may be occurring in the surface. A similar situation appears in the case of GJ 182 in terms of the recovered magnetic field topology. However, the field strength is much larger in this case (up to 400 G, \citealt{2008MNRAS.390..545D}), which is mostly connected with the rapid rotation, large latitudinal shear, and the nearly fully convective nature of this star. 

Regardless of the low $v\sin i$ of HD 1237, it is clear that a more complex field distribution is required to fit the observed spectra. Some weaker small-scale regions are recovered closer to the stellar equator. Despite their relative low strength compared to the total surface field, the contribution from these small-scale features to magnetically-related phenomena may be significantly higher. These features, in combination with strong mixed-polarity regions missed by ZDI, can influence the quiescent coronal emission \citep{2010MNRAS.404..101J}, the X-ray modulation \citep{2011MNRAS.410.2472A}, and the wind structure around, very active stars \citep{2014MNRAS.439.2122L}, and the Sun \citep{2013ApJ...764...32G}.

The large-scale field distribution appears similar at both of the observed epochs, confirming the observed chromospheric activity and longitudinal magnetic field behaviour. Using the final ZDI maps, and the derived synthetic Stokes V profiles, we were able to reproduce the variations of $B_{\ell}$ consistently (for both observed epochs). As an example, the maximum value of $B_\ell$ in Fig. \ref{fig_6}, coincides when the large positive polarity region of the radial field is located close to the limb (day 3). Similarly, the minimum value obtained for $B_\ell$ in Fig. \ref{fig_6}, results from the negative polarity region in the radial field map, this time located much closer to the disk centre. This corroborates the robustness of our ZDI magnetic maps for this system.  

As was mentioned in Sect. \ref{sec_properties}, HD 1237 has a planetary companion with a mean separation of 0.49 AU, and a projected mass of $M_{\rm p}\sin(i) = 3.37 \pm 0.14$ M$_{\jupiter}$. This exoplanet is relatively far out in comparison with other similar systems where ZDI maps of the host star are available (see \citealt{2013MNRAS.435.1451F} and references therein). However, the enhanced activity levels of HD 1237 and the relatively strong and complex surface magnetic field (in connection with its stellar wind), could impact significantly the conditions experienced by the exoplanet through its orbit. Previous parametric studies of this system have predicted a relatively large mass loss rate ($\sim$ 85 $\dot{M_{\odot}}$), which could even lead to magnetospheric radio emission from the exoplanet \citep{2005MNRAS.356.1053S}. This will be considered in a future study, incorporating the recovered ZDI magnetic field maps into a detailed 3D magneto-hydrodynamics (MHD) code (BATS-R-US, \citealt{1999JCoPh.154..284P, 2012JCoPh.231..870T}), originally developed and validated for the solar wind and corona (e.g. \citealt{2013ApJ...764...23S, 2014ApJ...782...81V}), and recently applied in the stellar context (e.g. \citealt{2014ApJ...783...55C, 2014ApJ...790...57C}).   

\section{Conclusions}\label{sec_conclusions}
\noindent We have reconstructed magnetic field maps of the young planet-hosting G-type star, HD 1237 using the technique of Zeeman Doppler Imaging. As part of this detailed spectro-polarimetric study we have found the following:

\begin{list}{$\circ$}{}
\item We find that assuming a Milne-Eddington approximation for the local line profile produces a better fit to the shape of the observed LSD profiles. This influences the amount magnetic structures that can be recovered in the magnetic field maps. 

\item We propose a robust method to define the ``stopping criterion'' in Zeeman Doppler Imaging techniques. This allows one to choose the optimum degree of fit, beyond which the model ceases to provide a good fit to the observed dataset due to the introduction of artifacts into the resulting image. We have successfully applied this to two different datasets, with vastly different degrees of phase sampling.

\item As part of our optimisation routines we recover a rotation period of 7.0 days. This is consistent with the measured chromospheric and coronal activity levels of the star ($\log(R^\prime_{HK}) = - 4.38 \pm 0.05$, this paper; $\log(L_{\rm X}) = 29.02 \pm 0.06$, \citealt{2008ApJ...687.1339K}).

\item The magnetic field reconstructions for HD 1237 are dominated by a band of strong uni-directional azimuthal field at high latitudes, accompanied by a complex multi-polar radial field distribution. The largest magnetic regions show field strengths of $\sim$\,90 and 50 G for the azimuthal and the radial field components, respectively. 

\item We note that the field topology recovered for HD 1237 is fully commensurate with studies of $\xi$ Boo A and GJ 182, the two other stars that are closest to it in the mass, period diagram \citep{2009ARA&A..47..333D}. Larger sample sizes are needed to confirm these trends. If confirmed, it will be possible to predict the global magnetic field topologies and therefore the extended environments of planet-hosting stars at various stages of their evolution based on these fundamental parameters. 

\item We will address the influence of the different magnetic scales on the coronal structure and wind properties of HD 1237 in a follow-up paper, using the ZDI maps presented here as boundary conditions (see \citealt{2010ApJ...721...80C, 2011ApJ...733...67C}). Possible star-planet interactions occurring in the system via transient (e.g. CME events, \citealt{2011ApJ...738..166C}) or quiescent phenomena (e.g planetary radio emission, \citealt{2005MNRAS.356.1053S, 2012MNRAS.423.3285V}) can be also considered in future work. 

\end{list}

\begin{acknowledgements}
\noindent Based on observations made with ESO Telescopes at the La Silla Paranal Observatory under programme ID 089.D-0138.  
\end{acknowledgements}


\bibliographystyle{aa}
\bibliography{Biblio}

\begin{thebibliography}{91}
\expandafter\ifx\csname natexlab\endcsname\relax\def\natexlab#1{#1}\fi

\bibitem[{{Anglada-Escud{\'e}} \& {Butler}(2012)}]{2012ApJS..200...15A}
{Anglada-Escud{\'e}}, G. \& {Butler}, R.~P. 2012, \apjs, 200, 15

\bibitem[{{Arzoumanian} {et~al.}(2011){Arzoumanian}, {Jardine}, {Donati},
  {Morin}, \& {Johnstone}}]{2011MNRAS.410.2472A}
{Arzoumanian}, D., {Jardine}, M., {Donati}, J.-F., {Morin}, J., \& {Johnstone},
  C. 2011, \mnras, 410, 2472

\bibitem[{{Bagnulo} {et~al.}(2009){Bagnulo}, {Landolfi}, {Landstreet}, {Landi
  Degl'Innocenti}, {Fossati}, \& {Sterzik}}]{2009PASP..121..993B}
{Bagnulo}, S., {Landolfi}, M., {Landstreet}, J.~D., {et~al.} 2009, \pasp, 121,
  993

\bibitem[{{Baliunas} {et~al.}(1995){Baliunas}, {Donahue}, {Soon}, {Horne},
  {Frazer}, {Woodard-Eklund}, {Bradford}, {Rao}, {Wilson}, {Zhang}, {Bennett},
  {Briggs}, {Carroll}, {Duncan}, {Figueroa}, {Lanning}, {Misch}, {Mueller},
  {Noyes}, {Poppe}, {Porter}, {Robinson}, {Russell}, {Shelton}, {Soyumer},
  {Vaughan}, \& {Whitney}}]{1995ApJ...438..269B}
{Baliunas}, S.~L., {Donahue}, R.~A., {Soon}, W.~H., {et~al.} 1995, \apj, 438,
  269

\bibitem[{{Barnes} {et~al.}(2004){Barnes}, {Lister}, {Hilditch}, \& {Collier
  Cameron}}]{2004MNRAS.348.1321B}
{Barnes}, J.~R., {Lister}, T.~A., {Hilditch}, R.~W., \& {Collier Cameron}, A.
  2004, \mnras, 348, 1321

\bibitem[{{Bastien} {et~al.}(2014){Bastien}, {Stassun}, {Pepper}, {Wright},
  {Aigrain}, {Basri}, {Johnson}, {Howard}, \&
  {Walkowicz}}]{2014AJ....147...29B}
{Bastien}, F.~A., {Stassun}, K.~G., {Pepper}, J., {et~al.} 2014, \aj, 147, 29

\bibitem[{{Boro Saikia} {et~al.}(2015){Boro Saikia}, {Jeffers}, {Petit},
  {Marsden}, {Morin}, \& {Folsom}}]{2015A&A...573A..17B}
{Boro Saikia}, S., {Jeffers}, S.~V., {Petit}, P., {et~al.} 2015, \aap, 573, A17

\bibitem[{{Brown} {et~al.}(2010){Brown}, {Browning}, {Brun}, {Miesch}, \&
  {Toomre}}]{2010ApJ...711..424B}
{Brown}, B.~P., {Browning}, M.~K., {Brun}, A.~S., {Miesch}, M.~S., \& {Toomre},
  J. 2010, \apj, 711, 424

\bibitem[{{Cohen} \& {Drake}(2014)}]{2014ApJ...783...55C}
{Cohen}, O. \& {Drake}, J.~J. 2014, \apj, 783, 55

\bibitem[{{Cohen} {et~al.}(2014){Cohen}, {Drake}, {Glocer}, {Garraffo},
  {Poppenhaeger}, {Bell}, {Ridley}, \& {Gombosi}}]{2014ApJ...790...57C}
{Cohen}, O., {Drake}, J.~J., {Glocer}, A., {et~al.} 2014, \apj, 790, 57

\bibitem[{{Cohen} {et~al.}(2010){Cohen}, {Drake}, {Kashyap}, {Hussain}, \&
  {Gombosi}}]{2010ApJ...721...80C}
{Cohen}, O., {Drake}, J.~J., {Kashyap}, V.~L., {Hussain}, G.~A.~J., \&
  {Gombosi}, T.~I. 2010, \apj, 721, 80

\bibitem[{{Cohen} {et~al.}(2011{\natexlab{a}}){Cohen}, {Kashyap}, {Drake},
  {Sokolov}, {Garraffo}, \& {Gombosi}}]{2011ApJ...733...67C}
{Cohen}, O., {Kashyap}, V.~L., {Drake}, J.~J., {et~al.} 2011{\natexlab{a}},
  \apj, 733, 67

\bibitem[{{Cohen} {et~al.}(2011{\natexlab{b}}){Cohen}, {Kashyap}, {Drake},
  {Sokolov}, \& {Gombosi}}]{2011ApJ...738..166C}
{Cohen}, O., {Kashyap}, V.~L., {Drake}, J.~J., {Sokolov}, I.~V., \& {Gombosi},
  T.~I. 2011{\natexlab{b}}, \apj, 738, 166

\bibitem[{{Collier Cameron}(1995)}]{1995MNRAS.275..534C}
{Collier Cameron}, A. 1995, \mnras, 275, 534

\bibitem[{{Daou} {et~al.}(2006){Daou}, {Johns-Krull}, \&
  {Valenti}}]{2006AJ....131..520D}
{Daou}, A.~G., {Johns-Krull}, C.~M., \& {Valenti}, J.~A. 2006, \aj, 131, 520

\bibitem[{{Donati}(2003)}]{2003ASPC..307...41D}
{Donati}, J.-F. 2003, in Astronomical Society of the Pacific Conference Series,
  Vol. 307, Solar Polarization, ed. J.~{Trujillo-Bueno} \& J.~{Sanchez
  Almeida}, 41

\bibitem[{{Donati} \& {Brown}(1997)}]{1997A&A...326.1135D}
{Donati}, J.-F. \& {Brown}, S.~F. 1997, \aap, 326, 1135

\bibitem[{{Donati} {et~al.}(2012){Donati}, {Gregory}, {Alencar}, {Hussain},
  {Bouvier}, {Dougados}, {Jardine}, {M{\'e}nard}, \&
  {Romanova}}]{2012MNRAS.425.2948D}
{Donati}, J.-F., {Gregory}, S.~G., {Alencar}, S.~H.~P., {et~al.} 2012, \mnras,
  425, 2948

\bibitem[{{Donati} {et~al.}(2014){Donati}, {H{\'e}brard}, {Hussain}, {Moutou},
  {Grankin}, {Boisse}, {Morin}, {Gregory}, {Vidotto}, {Bouvier}, {Alencar},
  {Delfosse}, {Doyon}, {Takami}, {Jardine}, {Fares}, {Cameron}, {M{\'e}nard},
  {Dougados}, {Herczeg}, \& {Matysse Collaboration}}]{2014MNRAS.444.3220D}
{Donati}, J.-F., {H{\'e}brard}, E., {Hussain}, G., {et~al.} 2014, \mnras, 444,
  3220

\bibitem[{{Donati} {et~al.}(2008{\natexlab{a}}){Donati}, {Jardine}, {Gregory},
  {Petit}, {Paletou}, {Bouvier}, {Dougados}, {M{\'e}nard}, {Collier Cameron},
  {Harries}, {Hussain}, {Unruh}, {Morin}, {Marsden}, {Manset}, {Auri{\`e}re},
  {Catala}, \& {Alecian}}]{2008MNRAS.386.1234D}
{Donati}, J.-F., {Jardine}, M.~M., {Gregory}, S.~G., {et~al.}
  2008{\natexlab{a}}, \mnras, 386, 1234

\bibitem[{{Donati} \& {Landstreet}(2009)}]{2009ARA&A..47..333D}
{Donati}, J.-F. \& {Landstreet}, J.~D. 2009, \araa, 47, 333

\bibitem[{{Donati} {et~al.}(2008{\natexlab{b}}){Donati}, {Morin}, {Petit},
  {Delfosse}, {Forveille}, {Auri{\`e}re}, {Cabanac}, {Dintrans}, {Fares},
  {Gastine}, {Jardine}, {Ligni{\`e}res}, {Paletou}, {Ramirez Velez}, \&
  {Th{\'e}ado}}]{2008MNRAS.390..545D}
{Donati}, J.-F., {Morin}, J., {Petit}, P., {et~al.} 2008{\natexlab{b}}, \mnras,
  390, 545

\bibitem[{{Donati} {et~al.}(1997){Donati}, {Semel}, {Carter}, {Rees}, \&
  {Collier Cameron}}]{1997MNRAS.291..658D}
{Donati}, J.-F., {Semel}, M., {Carter}, B.~D., {Rees}, D.~E., \& {Collier
  Cameron}, A. 1997, \mnras, 291, 658

\bibitem[{{Dumusque} {et~al.}(2014){Dumusque}, {Boisse}, \&
  {Santos}}]{2014ApJ...796..132D}
{Dumusque}, X., {Boisse}, I., \& {Santos}, N.~C. 2014, \apj, 796, 132

\bibitem[{{Dunstone} {et~al.}(2008){Dunstone}, {Hussain}, {Collier Cameron},
  {Marsden}, {Jardine}, {Stempels}, {Ramirez Velez}, \&
  {Donati}}]{2008MNRAS.387..481D}
{Dunstone}, N.~J., {Hussain}, G.~A.~J., {Collier Cameron}, A., {et~al.} 2008,
  \mnras, 387, 481

\bibitem[{{Eisenbeiss} {et~al.}(2013){Eisenbeiss}, {Ammler-von Eiff}, {Roell},
  {Mugrauer}, {Adam}, {Neuh{\"a}user}, {Schmidt}, \&
  {Bedalov}}]{2013A&A...556A..53E}
{Eisenbeiss}, T., {Ammler-von Eiff}, M., {Roell}, T., {et~al.} 2013, \aap, 556,
  A53

\bibitem[{{Ekenb{\"a}ck} {et~al.}(2010){Ekenb{\"a}ck}, {Holmstr{\"o}m}, {Wurz},
  {Grie{\ss}meier}, {Lammer}, {Selsis}, \& {Penz}}]{2010ApJ...709..670E}
{Ekenb{\"a}ck}, A., {Holmstr{\"o}m}, M., {Wurz}, P., {et~al.} 2010, \apj, 709,
  670

\bibitem[{{Fares}(2014)}]{2014IAUS..302..180F}
{Fares}, R. 2014, in IAU Symposium, Vol. 302, IAU Symposium, 180--189

\bibitem[{{Fares} {et~al.}(2013){Fares}, {Moutou}, {Donati}, {Catala},
  {Shkolnik}, {Jardine}, {Cameron}, \& {Deleuil}}]{2013MNRAS.435.1451F}
{Fares}, R., {Moutou}, C., {Donati}, J.-F., {et~al.} 2013, \mnras, 435, 1451

\bibitem[{{Folsom} {et~al.}(2013){Folsom}, {Petit}, {Bouvier}, {Donati}, \&
  {Morin}}]{2013arXiv1310.2073F}
{Folsom}, C.~P., {Petit}, P., {Bouvier}, J., {Donati}, J.-F., \& {Morin}, J.
  2013, ArXiv e-prints

\bibitem[{{Garraffo} {et~al.}(2013){Garraffo}, {Cohen}, {Drake}, \&
  {Downs}}]{2013ApJ...764...32G}
{Garraffo}, C., {Cohen}, O., {Drake}, J.~J., \& {Downs}, C. 2013, \apj, 764, 32

\bibitem[{{Ghezzi} {et~al.}(2010){Ghezzi}, {Cunha}, {Smith}, {de Ara{\'u}jo},
  {Schuler}, \& {de la Reza}}]{2010ApJ...720.1290G}
{Ghezzi}, L., {Cunha}, K., {Smith}, V.~V., {et~al.} 2010, \apj, 720, 1290

\bibitem[{{Gray} {et~al.}(1996){Gray}, {Baliunas}, {Lockwood}, \&
  {Skiff}}]{1996ApJ...465..945G}
{Gray}, D.~F., {Baliunas}, S.~L., {Lockwood}, G.~W., \& {Skiff}, B.~A. 1996,
  \apj, 465, 945

\bibitem[{{Hussain} {et~al.}(2009){Hussain}, {Collier Cameron}, {Jardine},
  {Dunstone}, {Ramirez Velez}, {Stempels}, {Donati}, {Semel}, {Aulanier},
  {Harries}, {Bouvier}, {Dougados}, {Ferreira}, {Carter}, \&
  {Lawson}}]{2009MNRAS.398..189H}
{Hussain}, G.~A.~J., {Collier Cameron}, A., {Jardine}, M.~M., {et~al.} 2009,
  \mnras, 398, 189

\bibitem[{{Hussain} {et~al.}(2000){Hussain}, {Donati}, {Collier Cameron}, \&
  {Barnes}}]{2000MNRAS.318..961H}
{Hussain}, G.~A.~J., {Donati}, J.-F., {Collier Cameron}, A., \& {Barnes}, J.~R.
  2000, \mnras, 318, 961

\bibitem[{{Janson} {et~al.}(2008){Janson}, {Reffert}, {Brandner}, {Henning},
  {Lenzen}, \& {Hippler}}]{2008A&A...488..771J}
{Janson}, M., {Reffert}, S., {Brandner}, W., {et~al.} 2008, \aap, 488, 771

\bibitem[{{Jeffers} {et~al.}(2014{\natexlab{a}}){Jeffers}, {Barnes}, {Jones},
  {Reiners}, {Pinfield}, \& {Marsden}}]{2014MNRAS.438.2717J}
{Jeffers}, S.~V., {Barnes}, J.~R., {Jones}, H.~R.~A., {et~al.}
  2014{\natexlab{a}}, \mnras, 438, 2717

\bibitem[{{Jeffers} {et~al.}(2014{\natexlab{b}}){Jeffers}, {Petit}, {Marsden},
  {Morin}, {Donati}, \& {Folsom}}]{2014A&A...569A..79J}
{Jeffers}, S.~V., {Petit}, P., {Marsden}, S.~C., {et~al.} 2014{\natexlab{b}},
  \aap, 569, A79

\bibitem[{{Jensen} {et~al.}(2012){Jensen}, {Redfield}, {Endl}, {Cochran},
  {Koesterke}, \& {Barman}}]{2012ApJ...751...86J}
{Jensen}, A.~G., {Redfield}, S., {Endl}, M., {et~al.} 2012, \apj, 751, 86

\bibitem[{{Johnstone} {et~al.}(2010){Johnstone}, {Jardine}, \&
  {Mackay}}]{2010MNRAS.404..101J}
{Johnstone}, C., {Jardine}, M., \& {Mackay}, D.~H. 2010, \mnras, 404, 101

\bibitem[{{Kashyap} {et~al.}(2008){Kashyap}, {Drake}, \&
  {Saar}}]{2008ApJ...687.1339K}
{Kashyap}, V.~L., {Drake}, J.~J., \& {Saar}, S.~H. 2008, \apj, 687, 1339

\bibitem[{{Kochukhov} {et~al.}(2010){Kochukhov}, {Makaganiuk}, \&
  {Piskunov}}]{2010A&A...524A...5K}
{Kochukhov}, O., {Makaganiuk}, V., \& {Piskunov}, N. 2010, \aap, 524, A5

\bibitem[{{Koen} {et~al.}(2010){Koen}, {Kilkenny}, {van Wyk}, \&
  {Marang}}]{2010MNRAS.403.1949K}
{Koen}, C., {Kilkenny}, D., {van Wyk}, F., \& {Marang}, F. 2010, \mnras, 403,
  1949

\bibitem[{{Kotov} {et~al.}(1998){Kotov}, {Scherrer}, {Howard}, \&
  {Haneychuk}}]{1998ApJS..116..103K}
{Kotov}, V.~A., {Scherrer}, P.~H., {Howard}, R.~F., \& {Haneychuk}, V.~I. 1998,
  \apjs, 116, 103

\bibitem[{{Kupka} {et~al.}(2000){Kupka}, {Ryabchikova}, {Piskunov}, {Stempels},
  \& {Weiss}}]{2000BaltA...9..590K}
{Kupka}, F.~G., {Ryabchikova}, T.~A., {Piskunov}, N.~E., {Stempels}, H.~C., \&
  {Weiss}, W.~W. 2000, Baltic Astronomy, 9, 590

\bibitem[{{Lang} {et~al.}(2014){Lang}, {Jardine}, {Morin}, {Donati}, {Jeffers},
  {Vidotto}, \& {Fares}}]{2014MNRAS.439.2122L}
{Lang}, P., {Jardine}, M., {Morin}, J., {et~al.} 2014, \mnras, 439, 2122

\bibitem[{{Linsky} {et~al.}(2010){Linsky}, {Yang}, {France}, {Froning},
  {Green}, {Stocke}, \& {Osterman}}]{2010ApJ...717.1291L}
{Linsky}, J.~L., {Yang}, H., {France}, K., {et~al.} 2010, \apj, 717, 1291

\bibitem[{{Liu} {et~al.}(2004){Liu}, {Matthews}, {Williams}, \&
  {Kalas}}]{2004ApJ...608..526L}
{Liu}, M.~C., {Matthews}, B.~C., {Williams}, J.~P., \& {Kalas}, P.~G. 2004,
  \apj, 608, 526

\bibitem[{{Lockwood} {et~al.}(2007){Lockwood}, {Skiff}, {Henry}, {Henry},
  {Radick}, {Baliunas}, {Donahue}, \& {Soon}}]{2007ApJS..171..260L}
{Lockwood}, G.~W., {Skiff}, B.~A., {Henry}, G.~W., {et~al.} 2007, \apjs, 171,
  260

\bibitem[{{Makaganiuk} {et~al.}(2011){Makaganiuk}, {Kochukhov}, {Piskunov},
  {Jeffers}, {Johns-Krull}, {Keller}, {Rodenhuis}, {Snik}, {Stempels}, \&
  {Valenti}}]{2011A&A...525A..97M}
{Makaganiuk}, V., {Kochukhov}, O., {Piskunov}, N., {et~al.} 2011, \aap, 525,
  A97

\bibitem[{{Mamajek} \& {Hillenbrand}(2008)}]{2008ApJ...687.1264M}
{Mamajek}, E.~E. \& {Hillenbrand}, L.~A. 2008, \apj, 687, 1264

\bibitem[{{Markwardt}(2009)}]{markwardt09}
{Markwardt}, C.~B. 2009, in Astronomical Society of the Pacific Conference
  Series, Vol. 411, Astronomical Data Analysis Software and Systems XVIII, ed.
  D.~A. {Bohlender}, D.~{Durand}, \& P.~{Dowler}, 251

\bibitem[{{Marsden} {et~al.}(2014){Marsden}, {Petit}, {Jeffers}, {Morin},
  {Fares}, {Reiners}, {do Nascimento}, {Auri{\`e}re}, {Bouvier}, {Carter},
  {Catala}, {Dintrans}, {Donati}, {Gastine}, {Jardine}, {Konstantinova-Antova},
  {Lanoux}, {Ligni{\`e}res}, {Morgenthaler}, {Ram{\`i}rez-V{\`e}lez},
  {Th{\'e}ado}, {Van Grootel}, \& {BCool Collaboration}}]{2014MNRAS.444.3517M}
{Marsden}, S.~C., {Petit}, P., {Jeffers}, S.~V., {et~al.} 2014, \mnras, 444,
  3517

\bibitem[{{Mayor} {et~al.}(2003){Mayor}, {Pepe}, {Queloz}, {Bouchy},
  {Rupprecht}, {Lo Curto}, {Avila}, {Benz}, {Bertaux}, {Bonfils}, {Dall},
  {Dekker}, {Delabre}, {Eckert}, {Fleury}, {Gilliotte}, {Gojak}, {Guzman},
  {Kohler}, {Lizon}, {Longinotti}, {Lovis}, {Megevand}, {Pasquini}, {Reyes},
  {Sivan}, {Sosnowska}, {Soto}, {Udry}, {van Kesteren}, {Weber}, \&
  {Weilenmann}}]{2003Msngr.114...20M}
{Mayor}, M., {Pepe}, F., {Queloz}, D., {et~al.} 2003, The Messenger, 114, 20

\bibitem[{{Middelkoop}(1982)}]{1982A&A...107...31M}
{Middelkoop}, F. 1982, \aap, 107, 31

\bibitem[{Mor{\' e}(1978)}]{more78}
Mor{\' e}, J. 1978, in Lecture Notes in Mathematics, Vol. 630, Numerical
  Analysis, ed. G.~Watson (Springer Berlin Heidelberg), 105--116

\bibitem[{{Morgenthaler} {et~al.}(2012){Morgenthaler}, {Petit}, {Saar},
  {Solanki}, {Morin}, {Marsden}, {Auri{\`e}re}, {Dintrans}, {Fares}, {Gastine},
  {Lanoux}, {Ligni{\`e}res}, {Paletou}, {Ram{\'{\i}}rez V{\'e}lez},
  {Th{\'e}ado}, \& {Van Grootel}}]{2012A&A...540A.138M}
{Morgenthaler}, A., {Petit}, P., {Saar}, S., {et~al.} 2012, \aap, 540, A138

\bibitem[{{Morin} {et~al.}(2008){Morin}, {Donati}, {Petit}, {Delfosse},
  {Forveille}, {Albert}, {Auri{\`e}re}, {Cabanac}, {Dintrans}, {Fares},
  {Gastine}, {Jardine}, {Ligni{\`e}res}, {Paletou}, {Ramirez Velez}, \&
  {Th{\'e}ado}}]{2008MNRAS.390..567M}
{Morin}, J., {Donati}, J.-F., {Petit}, P., {et~al.} 2008, \mnras, 390, 567

\bibitem[{{Naef} {et~al.}(2001){Naef}, {Mayor}, {Pepe}, {Queloz}, {Santos},
  {Udry}, \& {Burnet}}]{2001A&A...375..205N}
{Naef}, D., {Mayor}, M., {Pepe}, F., {et~al.} 2001, \aap, 375, 205

\bibitem[{{Neiner} {et~al.}(2012){Neiner}, {Grunhut}, {Petit}, {ud-Doula},
  {Wade}, {Landstreet}, {de Batz}, {Cochard}, {Guti{\'e}rrez-Soto}, \&
  {Huat}}]{2012MNRAS.426.2738N}
{Neiner}, C., {Grunhut}, J.~H., {Petit}, V., {et~al.} 2012, \mnras, 426, 2738

\bibitem[{{Noyes} {et~al.}(1984){Noyes}, {Hartmann}, {Baliunas}, {Duncan}, \&
  {Vaughan}}]{1984ApJ...279..763N}
{Noyes}, R.~W., {Hartmann}, L.~W., {Baliunas}, S.~L., {Duncan}, D.~K., \&
  {Vaughan}, A.~H. 1984, \apj, 279, 763

\bibitem[{{Peres} {et~al.}(2000){Peres}, {Orlando}, {Reale}, {Rosner}, \&
  {Hudson}}]{2000ApJ...528..537P}
{Peres}, G., {Orlando}, S., {Reale}, F., {Rosner}, R., \& {Hudson}, H. 2000,
  \apj, 528, 537

\bibitem[{{Petit} {et~al.}(2008){Petit}, {Dintrans}, {Solanki}, {Donati},
  {Auri{\`e}re}, {Ligni{\`e}res}, {Morin}, {Paletou}, {Ramirez Velez},
  {Catala}, \& {Fares}}]{2008MNRAS.388...80P}
{Petit}, P., {Dintrans}, B., {Solanki}, S.~K., {et~al.} 2008, \mnras, 388, 80

\bibitem[{{Petit} {et~al.}(2002){Petit}, {Donati}, \& {Collier
  Cameron}}]{2002MNRAS.334..374P}
{Petit}, P., {Donati}, J.-F., \& {Collier Cameron}, A. 2002, \mnras, 334, 374

\bibitem[{{Piskunov} \& {Kochukhov}(2002)}]{2002A&A...381..736P}
{Piskunov}, N. \& {Kochukhov}, O. 2002, \aap, 381, 736

\bibitem[{{Piskunov} {et~al.}(2011){Piskunov}, {Snik}, {Dolgopolov},
  {Kochukhov}, {Rodenhuis}, {Valenti}, {Jeffers}, {Makaganiuk}, {Johns-Krull},
  {Stempels}, \& {Keller}}]{2011Msngr.143....7P}
{Piskunov}, N., {Snik}, F., {Dolgopolov}, A., {et~al.} 2011, The Messenger,
  143, 7

\bibitem[{{Piskunov} \& {Valenti}(2002)}]{2002A&A...385.1095P}
{Piskunov}, N.~E. \& {Valenti}, J.~A. 2002, \aap, 385, 1095

\bibitem[{{Poppenhaeger} {et~al.}(2011){Poppenhaeger}, {Robrade}, \&
  {Schmitt}}]{2011A&A...529C...1P}
{Poppenhaeger}, K., {Robrade}, J., \& {Schmitt}, J.~H.~M.~M. 2011, \aap, 529,
  C1

\bibitem[{{Powell} {et~al.}(1999){Powell}, {Roe}, {Linde}, {Gombosi}, \& {De
  Zeeuw}}]{1999JCoPh.154..284P}
{Powell}, K.~G., {Roe}, P.~L., {Linde}, T.~J., {Gombosi}, T.~I., \& {De Zeeuw},
  D.~L. 1999, Journal of Computational Physics, 154, 284

\bibitem[{{Saffe} {et~al.}(2005){Saffe}, {G{\'o}mez}, \&
  {Chavero}}]{2005A&A...443..609S}
{Saffe}, C., {G{\'o}mez}, M., \& {Chavero}, C. 2005, \aap, 443, 609

\bibitem[{{Santos} {et~al.}(2000){Santos}, {Mayor}, {Naef}, {Pepe}, {Queloz},
  {Udry}, \& {Blecha}}]{2000A&A...361..265S}
{Santos}, N.~C., {Mayor}, M., {Naef}, D., {et~al.} 2000, \aap, 361, 265

\bibitem[{{Sanz-Forcada} {et~al.}(2011){Sanz-Forcada}, {Micela}, {Ribas},
  {Pollock}, {Eiroa}, {Velasco}, {Solano}, \&
  {Garc{\'{\i}}a-{\'A}lvarez}}]{2011A&A...532A...6S}
{Sanz-Forcada}, J., {Micela}, G., {Ribas}, I., {et~al.} 2011, \aap, 532, A6

\bibitem[{{Schmitt} \& {Liefke}(2004)}]{2004A&A...417..651S}
{Schmitt}, J.~H.~M.~M. \& {Liefke}, C. 2004, \aap, 417, 651

\bibitem[{{Schr{\"o}der} {et~al.}(2009){Schr{\"o}der}, {Reiners}, \&
  {Schmitt}}]{2009A&A...493.1099S}
{Schr{\"o}der}, C., {Reiners}, A., \& {Schmitt}, J.~H.~M.~M. 2009, \aap, 493,
  1099

\bibitem[{{Semel}(1989)}]{1989A&A...225..456S}
{Semel}, M. 1989, \aap, 225, 456

\bibitem[{{Shibata} \& {Magara}(2011)}]{2011LRSP....8....6S}
{Shibata}, K. \& {Magara}, T. 2011, Living Reviews in Solar Physics, 8, 6

\bibitem[{{Sing}(2010)}]{2010A&A...510A..21S}
{Sing}, D.~K. 2010, \aap, 510, A21

\bibitem[{{Sokolov} {et~al.}(2013){Sokolov}, {van der Holst}, {Oran}, {Downs},
  {Roussev}, {Jin}, {Manchester}, {Evans}, \& {Gombosi}}]{2013ApJ...764...23S}
{Sokolov}, I.~V., {van der Holst}, B., {Oran}, R., {et~al.} 2013, \apj, 764, 23

\bibitem[{Sonka {et~al.}(2007)Sonka, Hlavac, \& Boyle}]{Sonka:2007:IPA:1210103}
Sonka, M., Hlavac, V., \& Boyle, R. 2007, Image Processing, Analysis, and
  Machine Vision (Thomson-Engineering)

\bibitem[{{Stevens}(2005)}]{2005MNRAS.356.1053S}
{Stevens}, I.~R. 2005, \mnras, 356, 1053

\bibitem[{{Torres} {et~al.}(2006){Torres}, {Quast}, {da Silva}, {de La Reza},
  {Melo}, \& {Sterzik}}]{2006A&A...460..695T}
{Torres}, C.~A.~O., {Quast}, G.~R., {da Silva}, L., {et~al.} 2006, \aap, 460,
  695

\bibitem[{{T{\'o}th} {et~al.}(2012){T{\'o}th}, {van der Holst}, {Sokolov}, {De
  Zeeuw}, {Gombosi}, {Fang}, {Manchester}, {Meng}, {Najib}, {Powell}, {Stout},
  {Glocer}, {Ma}, \& {Opher}}]{2012JCoPh.231..870T}
{T{\'o}th}, G., {van der Holst}, B., {Sokolov}, I.~V., {et~al.} 2012, Journal
  of Computational Physics, 231, 870

\bibitem[{{van der Holst} {et~al.}(2014){van der Holst}, {Sokolov}, {Meng},
  {Jin}, {Manchester}, {T{\'o}th}, \& {Gombosi}}]{2014ApJ...782...81V}
{van der Holst}, B., {Sokolov}, I.~V., {Meng}, X., {et~al.} 2014, \apj, 782, 81

\bibitem[{{Vidal-Madjar} {et~al.}(2003){Vidal-Madjar}, {Lecavelier des Etangs},
  {D{\'e}sert}, {Ballester}, {Ferlet}, {H{\'e}brard}, \&
  {Mayor}}]{2003Natur.422..143V}
{Vidal-Madjar}, A., {Lecavelier des Etangs}, A., {D{\'e}sert}, J.-M., {et~al.}
  2003, \nat, 422, 143

\bibitem[{{Vidotto} {et~al.}(2012){Vidotto}, {Fares}, {Jardine}, {Donati},
  {Opher}, {Moutou}, {Catala}, \& {Gombosi}}]{2012MNRAS.423.3285V}
{Vidotto}, A.~A., {Fares}, R., {Jardine}, M., {et~al.} 2012, \mnras, 423, 3285

\bibitem[{{Vidotto} {et~al.}(2013){Vidotto}, {Jardine}, {Morin}, {Donati},
  {Lang}, \& {Russell}}]{2013A&A...557A..67V}
{Vidotto}, A.~A., {Jardine}, M., {Morin}, J., {et~al.} 2013, \aap, 557, A67

\bibitem[{{Vogt} {et~al.}(1987){Vogt}, {Penrod}, \&
  {Hatzes}}]{1987ApJ...321..496V}
{Vogt}, S.~S., {Penrod}, G.~D., \& {Hatzes}, A.~P. 1987, \apj, 321, 496

\bibitem[{{Watson} {et~al.}(2010){Watson}, {Littlefair}, {Collier Cameron},
  {Dhillon}, \& {Simpson}}]{2010MNRAS.408.1606W}
{Watson}, C.~A., {Littlefair}, S.~P., {Collier Cameron}, A., {Dhillon}, V.~S.,
  \& {Simpson}, E.~K. 2010, \mnras, 408, 1606

\bibitem[{{Wood}(2004)}]{2004LRSP....1....2W}
{Wood}, B.~E. 2004, Living Reviews in Solar Physics, 1, 2

\bibitem[{{Wood} \& {Linsky}(2010)}]{2010ApJ...717.1279W}
{Wood}, B.~E. \& {Linsky}, J.~L. 2010, \apj, 717, 1279

\bibitem[{{Zechmeister} \& {K{\"u}rster}(2009)}]{2009A&A...496..577Z}
{Zechmeister}, M. \& {K{\"u}rster}, M. 2009, \aap, 496, 577

\end{thebibliography}
\begin{appendix}
\section{Second Epoch Dataset (2012 Dec)} \label{app_1}

\begin{figure}[!ht]
\centering		
\includegraphics[trim=2cm 2.1cm 3.0cm 0cm, clip=true, width=\hsize]{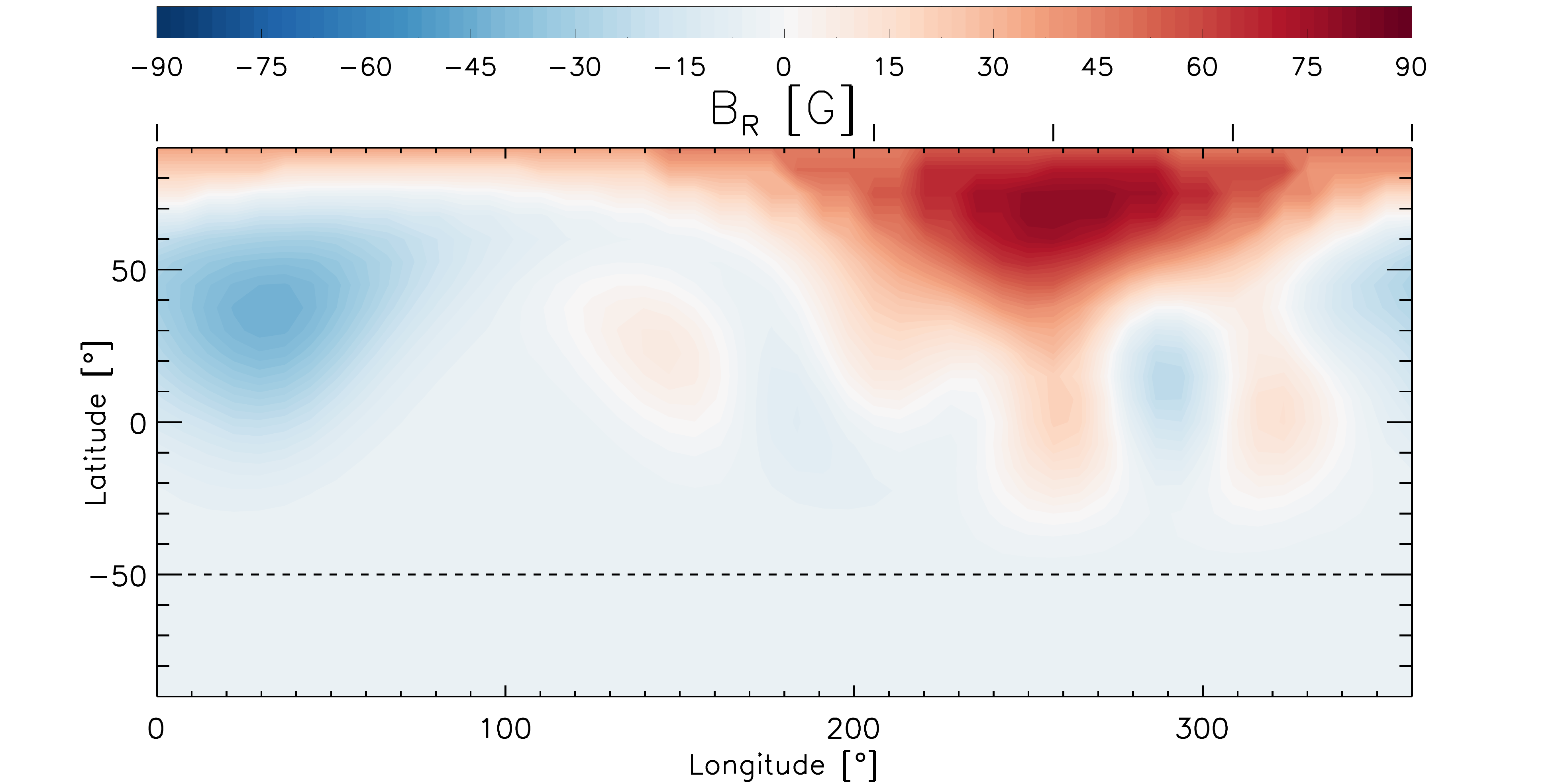} 
\includegraphics[trim=2cm 2.1cm 3.0cm 2cm, clip=true, width=\hsize]{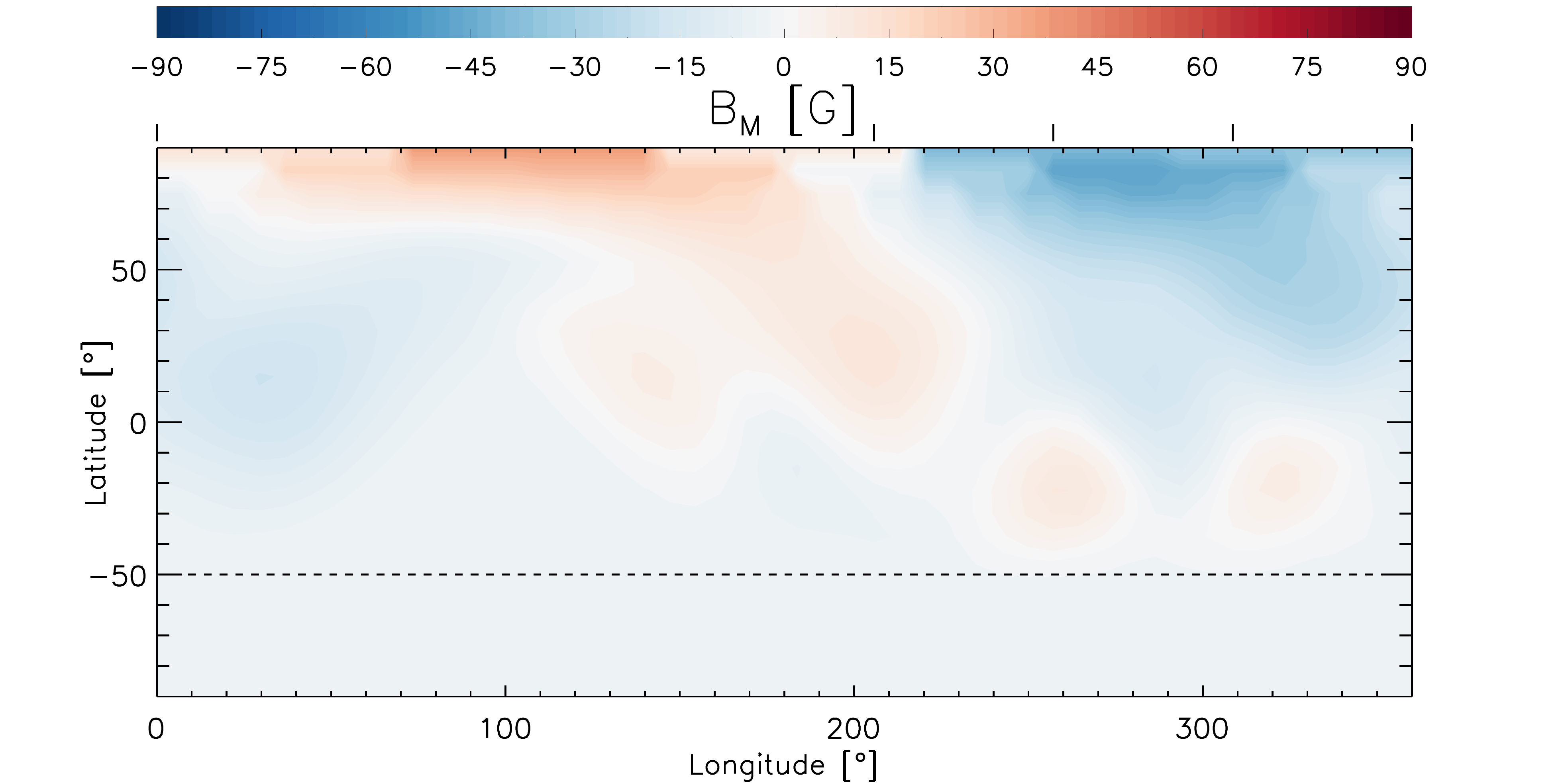} 
\includegraphics[trim=2cm 0cm 3.0cm 2cm, clip=true, width=\hsize]{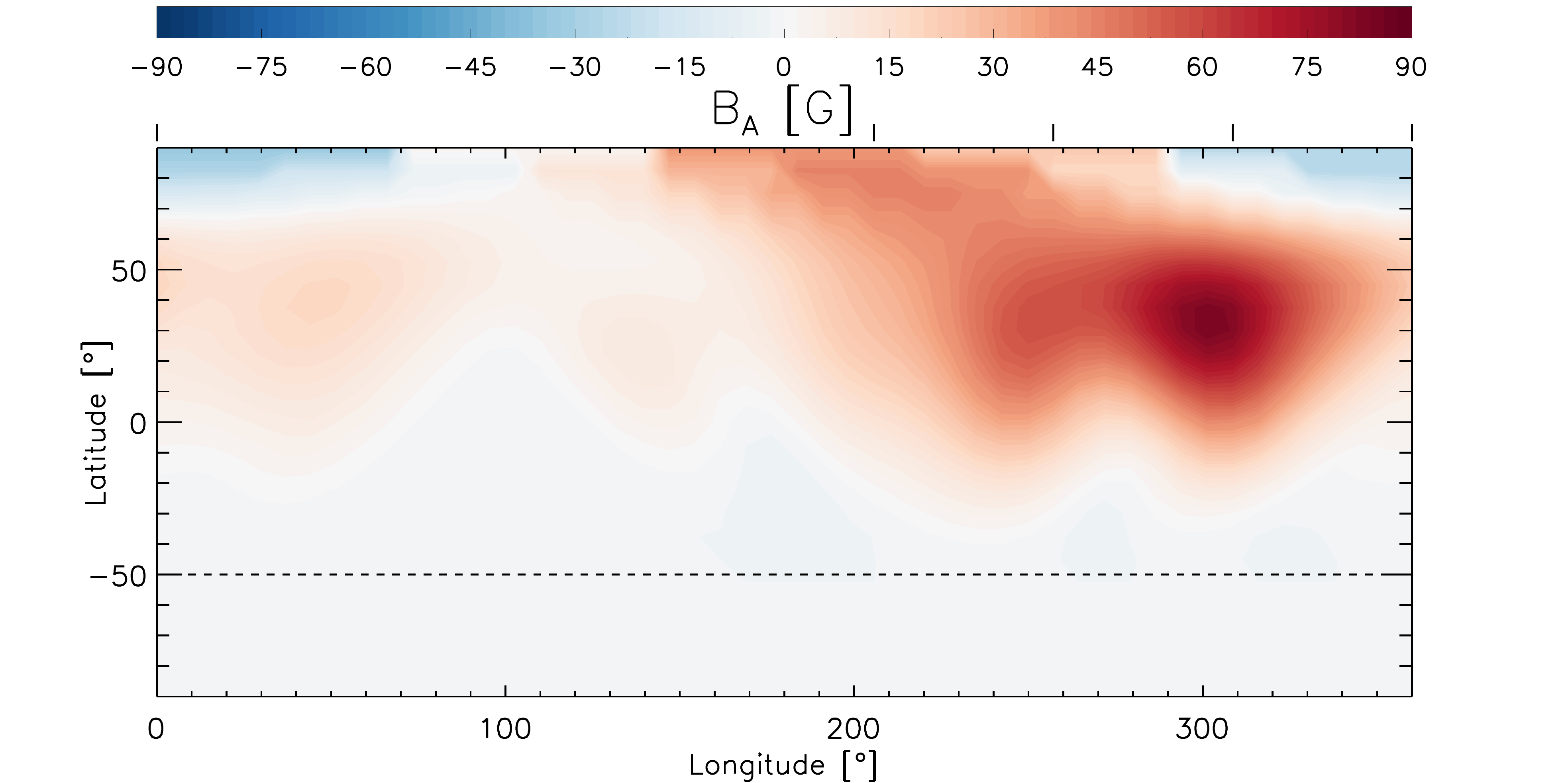} 
\includegraphics[trim=2.0cm 0.5cm 1.0cm 0.2cm, clip=true, width=\hsize]{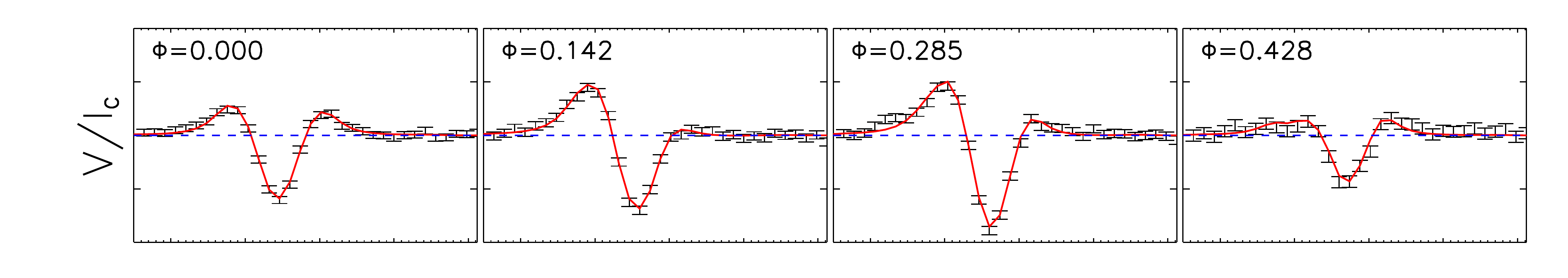} 
\caption{Results of the ZDI analysis for the second epoch observations of HD 1237 using the Milne-Eddington line profile. See caption of Fig. \ref{fig_8} for more details. In this case the maps fit the spectro-polarimetric data to an optimal reduced $\chi^2 = 0.6$. In this case, $\Phi = 0.0$ is assigned to the observations acquired at HJD 2456265.0.}
\label{fig_app}
\end{figure}

\noindent We consider only the Milne-Eddington line profile for the 2012 Dec dataset. Figure \ref{fig_app} contains the recovered ZDI maps for this epoch with the corresponding synthetic Stokes V profiles. Despite having a lower phase coverage in this case, we were able to recover robust magnetic field maps, fitting the spectro-polarimetric data up to an optimal reduced $\chi^2 = 0.6$ (see Sect. \ref{sec_criteria}). This small value of reduced $\chi^2$ results as a consequence of the fewer constraints available for this dataset. 

\noindent The field distribution clearly resembles the one obtained for the July dataset, with a large contribution from the azimuthal and radial components to the total field. The main large-scale magnetic features are preserved between both observed epochs. This is consistent with the behaviour shown by the activity indicators (Sect. \ref{sec_S_index}, Fig. \ref{fig_4}) and the longitudinal magnetic field (Sect. \ref{B_los}, Fig. \ref{fig_6}) in the entire dataset. Some of the smaller-scale structure is not recovered and the ring of azimuthal field is not clear in this case. These changes are expected due to the number of observations and phase coverage in this epoch (e.g. \citealt{1997A&A...326.1135D}). The slight shift in longitude is due to the initial phase selection in this case, where $\Phi = 0.0$ is assigned to the observations acquired at HJD 2456265.0 (Table \ref{tab_2}).

\end{appendix}

\end{document}